\RequirePackage{silence}
\documentclass[fleqn,usenatbib,useAMS]{mnras}
\usepackage{graphicx}
\usepackage{amsmath}
\usepackage{amsfonts}
\usepackage{float}
\usepackage{bm}
\usepackage[perpage,symbol*]{footmisc}
\usepackage{ae,aecompl}
\usepackage{array}
\usepackage{soul}
\usepackage{mathtools}
\usepackage{multirow}

\newcommand{\appropto}{\mathrel{\vcenter{
  \offinterlineskip\halign{\hfil$##$\cr 
    \propto\cr\noalign{\kern2pt}\sim\cr\noalign{\kern-2pt}}}}}

\newcommand{\ssim}{\,{\sim}\,} 

\DeclareRobustCommand{\perthousand}{%
  \ifmmode
    \text{\textperthousand}%
  \else
    \textperthousand
  \fi}

\title[A Plane Of High Velocity Local Group Galaxies]{A Plane of High Velocity Galaxies Across the Local Group} 

\author[Indranil Banik \& Hongsheng Zhao]{Indranil Banik$^{1}$\thanks{Email: \href{mailto:ib45@st-andrews.ac.uk}{ib45@st-andrews.ac.uk} (Indranil Banik)\newline $~~~~~~~~~~~~~~$ \href{mailto:hz4@st-andrews.ac.uk}{hz4@st-andrews.ac.uk} (Hongsheng Zhao)}, Hongsheng Zhao$^{1}$\\
$^{1}$Scottish Universities Physics Alliance, University of St Andrews, North Haugh, St Andrews, Fife, KY16 9SS, UK}

\pubyear{2017}
\begin{document}
\label{firstpage}
\pagerange{\pageref{firstpage}--\pageref{lastpage}}

\maketitle

\begin{abstract}

We recently showed that several Local Group (LG) galaxies have much higher radial velocities (RVs) than predicted by a 3D dynamical model of the standard cosmological paradigm. Here, we show that 6 of these 7 galaxies define a thin plane with root mean square thickness of only 101 kpc despite a widest extent of nearly 3 Mpc, much larger than the conventional virial radius of the Milky Way (MW) or M31. This plane passes within ${\ssim 70}$ kpc of the MW-M31 barycentre and is oriented so the MW-M31 line is inclined by $16^\circ$ to it.

We develop a toy model to constrain the scenario whereby a past MW-M31 flyby in Modified Newtonian Dynamics (MOND) forms tidal dwarf galaxies that settle into the recently discovered planes of satellites around the MW and M31. The scenario is viable only for a particular MW-M31 orbital plane. This roughly coincides with the plane of LG dwarfs with anomalously high RVs.

Using a restricted $N$-body simulation of the LG in MOND, we show how the once fast-moving MW and M31 gravitationally slingshot test particles outwards at high speeds. The most distant such particles preferentially lie within the MW-M31 orbital plane, probably because the particles ending up with the highest RVs are those flung out almost parallel to the motion of the perturber. This suggests a dynamical reason for our finding of a similar trend in the real LG, something not easily explained as a chance alignment of galaxies with an isotropic or mildly flattened distribution (probability $= {0.0015}$). 

\end{abstract}


\section{Introduction}
\label{Introduction}

The standard cosmological paradigm ($\Lambda$CDM) faces several challenges in the relatively well-observed Local Group (LG). In particular, the satellite systems of its two major galaxies $-$ the Milky Way (MW) and Andromeda (M31) $-$ both exhibit an unusual degree of anisotropy. For the MW, this has been suspected for several decades \citep{Lynden_Bell_1976, Lynden_Bell_1982}, though recent observations have greatly clarified the situation and its apparent tension with $\Lambda$CDM \citep{Kroupa_2005}. Proper motion measurements show that most of its satellites co-rotate within a well-defined plane \citep{Kroupa_2013}. Moreover, recently discovered ultra-faint satellites, globular clusters and tidal streams independently prefer a similarly oriented plane \citep{Pawlowski_2012}. Although some flattening is expected in $\Lambda$CDM \citep[e.g.][]{Butsky_2016}, it remains difficult to explain the very small thickness of the MW satellite system and its coherent rotation \citep{Pawlowski_2015}.

An analogous situation was suspected around M31 \citep{Metz_2007, Metz_2009}. This has recently been confirmed by \citet{Ibata_2013} using the Pan-Andromeda Archaeological Survey \citep{PANDAS}. Despite its greater distance, the detection of this highly anisotropic system is rather secure because it is almost edge-on as viewed from our perspective. A redshift gradient across it strongly suggests that it too is co-rotating \citep{Ibata_2013}. Like the MW satellite system, it is difficult for $\Lambda$CDM to explain the observed properties of the M31 satellite system \citep{Ibata_2014}.

Beyond the LG, satellite planes likely also exist around Centaurus A \citep{Muller_2016} and M81 \citep{Chiboucas_2013}. These discoveries may be related to a vast plane of dwarf galaxies recently found near M101 \citep{Muller_2017}. This does not consist solely of satellite galaxies because it extends over 3 Mpc. Such a structure appears difficult to find in cosmological simulations of the $\Lambda$CDM paradigm \citep[][Figure 8]{Gonzalez_2010}.

We recently uncovered another potential problem relating to the dynamics of non-satellite LG dwarf galaxies at distances of ${\sim 1 - 3}$ Mpc \citep{Banik_Zhao_2016}. This was based on a timing argument analysis of the LG \citep{Kahn_Woltjer_1959, Einasto_1982} extended to include test particles representing LG dwarfs. Following on from previous spherically symmetric dynamical models \citep{Sandage_1986, Jorge_2014}, we constructed an axisymmetric model of the LG consistent with the almost radial MW-M31 orbit \citep{Van_der_Marel_2012} and the close alignment of Centaurus A with this line \citep{Ma_1998}. Treating LG dwarfs as test particles in the gravitational field of these three massive moving objects, we investigated a wide range of model parameters using a full grid search. None of the models produced a good fit, even when we made reasonable allowance for inaccuracies in our model as a representation of $\Lambda$CDM based on the scatter about the Hubble flow in detailed $N$-body simulations of it \citep{Aragon_Calvo_2011}. This is because several LG dwarfs have Galactocentric Radial Velocities (GRVs) much higher than expected in our best-fitting model, though the opposite was rarely the case \citep[][Figure 9]{Banik_Zhao_2016}. We found that this should remain true even when certain factors beyond the model are included, in particular the Large Magellanic Cloud and the Great Attractor (GA).

We borrowed an algorithm described in \citet{Shaya_2011} to test whether this remains the case when using a three-dimensional (3D) model of the LG. The typical mismatch between observed and predicted GRVs in the best-fitting model is actually slightly higher than in the 2D case, with a clear tendency persisting for faster outward motion than expected \citep[][Figures 7 and 9]{Banik_Zhao_2017}. These results are similar to those obtained by \citet{Peebles_2017} using a similar algorithm. Despite a very different method to \citet{Banik_Zhao_2016}, the conclusions remain broadly similar.

Beyond the LG, another puzzling observation in a $\Lambda$CDM context is the remarkably tight correlation between the internal accelerations within galaxies (typically inferred from their rotation curves) and the prediction of Newtonian gravity applied to the distribution of their luminous matter \citep[e.g.][and references therein]{Famaey_McGaugh_2012}. This `radial acceleration relation' (RAR) is a generalisation of the baryonic Tully-Fisher relation \citep[e.g.][]{McGaugh_2015} which only considers the flat outer part of galaxy rotation curves and their total baryonic masses (Equation \ref{BTFR}), whereas the RAR considers all radii with accurate data.

The RAR has recently been confirmed and further tightened based on near-infrared photometry taken by the Spitzer Space Telescope \citep{SPARC}, considering only the most reliable rotation curves \citep[as described in its Section 3.2.2 and also in][]{Swaters_2009} and taking advantage of reduced variability in stellar mass-to-light ratios at these wavelengths \citep{Bell_de_Jong_2001, Norris_2016}. These improvements reveal that the RAR holds with very little scatter over ${\ssim 5}$ orders of magnitude in luminosity and a similar range of surface brightness \citep{McGaugh_Lelli_2016}.

In addition to disk galaxies, the RAR also seems to hold for ellipticals, whose internal forces can sometimes be measured accurately due to the presence of a thin rotation-supported gas disk \citep{Heijer_2015}. As well as these massive ellipticals, the RAR also works well in galaxies as faint as the satellites of M31 \citep{McGaugh_2013}. For a recent overview of how well the RAR works in several different types of galaxy across the Hubble sequence, we refer the reader to \citet{Lelli_2017}.

The RAR is either a fundamental consequence of natural law or an emergent property of galaxies relating their baryonic and dark matter distributions. The latter approach is taken by $\Lambda$CDM, a paradigm in which a relation of this sort is expected because lower mass dark matter halos have shallower gravitational potential wells. This should make it easier for baryons to be ejected via energetic processes like supernova feedback. Still, the tightness of the observed RAR is difficult to explain in this way \citep{Desmond_2017}. Some attempts have been made to do so \citep[e.g.][]{Keller_Wadsley_2017}, but so far these have investigated only a very small range of galaxy masses and types. In these limited circumstances, there does seem to be a tight correlation of the sort observed. However, a closer look reveals that several other aspects of the simulations are inconsistent with observations \citep{Milgrom_2016}. For example, the rotation curve amplitudes are significantly overestimated in the central regions \citep[][Figure 4]{Keller_Wadsley_2016}.

Unlike $\Lambda$CDM, Modified Newtonian Dynamics \citep[MOND,][]{Milgrom_1983} is predicated on the assumption that the RAR is fundamental and not due to galaxies being surrounded by dark matter halos. The dynamical effect of these halos is instead provided by a revised law of gravity arising from an acceleration-dependent modification to the Poisson equation of Newtonian gravity \citep{Bekenstein_Milgrom_1984, QUMOND}. In spherical symmetry, the gravitational field strength $g$ at distance $r$ from an isolated point mass $M$ transitions from the usual inverse square law at short range to
\begin{eqnarray}
	g ~=~ \frac{\sqrt{GMa_{_0}}}{r} ~~~\text{for } ~r \gg \sqrt{\frac{GM}{a_{_0}}}
	\label{Deep_MOND_limit}
\end{eqnarray}

Here, $a_{_0}$ is a fundamental acceleration scale of nature. Empirically, $a_{_0} \approx 1.2 \times {10}^{-10}$ m/s$^2$ to match galaxy rotation curves \citep{McGaugh_2011}. Remarkably, this is similar to the acceleration at which the energy density in a classical gravitational field becomes comparable to the dark energy density $u_{_\Lambda} = \rho_{_\Lambda} c^2$ implied by the accelerating expansion of the Universe \citep{Riess_1998}. Thus,
\begin{eqnarray}
	\frac{g^2}{8\rm{\pi}G} ~<~ u_{_\Lambda} ~~\Leftrightarrow~~ g ~\la~ 2\rm{\pi}a_{_0}
	\label{MOND_quantum_link}
\end{eqnarray}

This suggests that MOND may arise from quantum gravity effects \citep[e.g.][]{Milgrom_1999, Pazy_2013, Verlinde_2016, Smolin_2017}. Regardless of its underlying microphysical explanation, MOND can explain the Tully-Fisher relation \citep{Tully_Fisher_1977} as a specific example of the RAR by equating the gravitational field strength given by Equation \ref{Deep_MOND_limit} with the centripetal acceleration $\frac{v^2}{r}$ required to maintain a circular orbit. In the low-acceleration outskirts of galaxies beyond the extent of most of their visible mass\footnote{where a point mass approximation to the galaxy should be valid}, this predicts a flat rotation curve with amplitude
\begin{eqnarray}
	v_{_f} ~=~ \sqrt[4]{GMa_{_0}}
	\label{BTFR}
\end{eqnarray}


Although this is one of the more widely known consequences of MOND, the theory does much more than this and more even than the RAR, its prediction in isolated systems. For the recently discovered Crater 2 satellite of the MW \citep{Torrealba_2016}, it predicted the velocity dispersion to be a tiny 2.1 km/s \citep{McGaugh_2016}, partly due to a unique effect in MOND whereby its self-gravity is weakened by the external gravitational field of the nearby MW \citep[e.g.][]{Banik_Zhao_2015}. This was recently confirmed by observations, which are in tension with a naive application of the RAR but not a more rigorous treatment of MOND \citep{Caldwell_2017}. This external field effect hardly matters for calculating the rotation curve of the MW but is crucial to its escape velocity, measurements of which can be fit reasonably well in MOND \citep{Banik_2017_escape}. 

A crucial ingredient for the RAR is the strength of the gravitational field in the outskirts of galaxies. These are impossible to measure directly and can only be estimated from rotation curves. Gravitational lensing provides an independent way to check these estimates in a statistical sense. One such attempt was the Canada-France-Hawaii Telescope Lensing Survey \citep{Brimioulle_2013}. Stacked data from it shows that MOND can predict the correct amplitude of weak gravitational lensing by spiral and elliptical galaxies, using Equation \ref{Deep_MOND_limit} with the same value for $a_{_0}$ as that required to match disk galaxy rotation curves \citep{Milgrom_2013}. Thus, weak lensing and rotation curve measurements broadly agree on the strength of gravity in the outskirts of galaxies.\footnote{Although MOND is a non-relativistic theory, all attempts to generalise it to the relativistic case imply that the non-relativistic gravitational field determines light deflection in the same way as in General Relativity \citep[][Section 2]{Milgrom_2013}.}

As well as affecting forces within a galaxy, MOND also affects forces between them. In the LG, this implies a much stronger MW-M31 mutual attraction than $\Lambda$CDM. Combined with the almost radial nature of their relative motion \citep{Van_der_Marel_2012}, this means that they must have undergone a close encounter $\ssim 9 \pm 2$ Gyr ago \citep{Zhao_2013}. This could have led to the formation of a thin tidal tail which later condensed into satellite galaxies of the MW and M31, a phenomenon which seems to occur in some observed galactic interactions \citep{Mirabel_1992} and in MOND simulations of them \citep{Tiret_2008}.The formation mechanism of these tidal dwarf galaxies would lead to them lying close to a plane and co-rotating within that plane \citep{Wetzstein_2007}, though a small fraction might well end up counter-rotating \citep{Pawlowski_2011}. Some could even become unbound from both the MW and M31, instead flying away from the LG at high speed. This is possible once the effect of dark energy is considered as its repulsive effect rises with distance, unlike the gravitational field from a finite distribution of matter (Equation \ref{Test_particle_acceleration}).


A past MW-M31 interaction might also have formed the thick disk of the MW \citep{Gilmore_1983}, a structure which seems to have formed fairly rapidly from its thin disk ${9 \pm 1}$ Gyr ago \citep{Quillen_2001}. More recent investigations suggest a fairly rapid formation timescale \citep{Hayden_2015} and an associated burst of star formation \citep[][Figure 2]{Snaith_2014}. The disk heating which likely formed the Galactic thick disk appears to have been stronger in the outer parts of the MW, characteristic of a tidal effect \citep{Banik_2014}. This may be why the thick disk of the MW has a larger scale length than its thin disk \citep{Juric_2008, Jayaraman_2013}.

The high MW-M31 relative velocity around the time of their encounter \citep[${\ssim 600}$ km/s,][]{Zhao_2013} suggests that they could well have flung out several LG dwarfs at high speed in what would essentially have been 3-body gravitational interactions \citep{Banik_Zhao_2016}. The main objective of the present contribution is to test certain aspects of this scenario. In Section \ref{Geometry}, we extract some of its likely consequences based on a toy model of a past MW-M31 flyby encounter. Linking this to the observed geometry of the LG gives a constraint on the MW-M31 orbital plane. In Section \ref{MOND_simulation}, we use this in a more detailed MOND simulation of the LG incorporating several hundred thousand test particles affected by the gravity of the MW and M31, which undergo a close (14.17 kpc) flyby 6.59 Gyr after the Big Bang. As expected, some particles are flung out at high radial velocities after passing close to the spacetime location of this event. The particles flung out to the greatest distances have orbital angular momenta aligning rather closely with that of the MW-M31 orbit (Figure \ref{Coplanar_fraction_results}) and lie rather close to the MW-M31 orbital plane (Figure \ref{z_distribution}). This is probably because such particles were ejected almost parallel to the motion of the perturbing body in order to gain the most energy from it.

In Section \ref{Model_refinements}, we refine our previous $\Lambda$CDM model of the LG in 3D \citep{Banik_Zhao_2017} to help us better select LG galaxies whose kinematics suggest that they were flung out in this way. We quantify the spatial anisotropy of these galaxies in Section \ref{Method}. Here, we use our MOND-based simulation to identify three further properties that we expect of these high-velocity galaxies (HVGs). In Section \ref{Results}, we quantify how likely it is for a random distribution of HVGs to match these properties as well as the observed HVG system. In Section \ref{Discussion}, we discuss our analysis in light of previous works and consider some possibilities for explaining our results within MOND (Section \ref{Discussion_MOND}) and $\Lambda$CDM (Section \ref{Discussion_LCDM}). Our conclusions are given in Section \ref{Conclusions}.

\section{Geometry of a past MW-M31 Flyby}
\label{Geometry}
\subsection{Orientation of the M31 disk}
\label{M31_disk_spin_determination}

\begin{figure}
	\centering 
	\begin{minipage}[t]{0.19\textwidth}
		\vspace{5pt}
		\includegraphics [width=\textwidth] {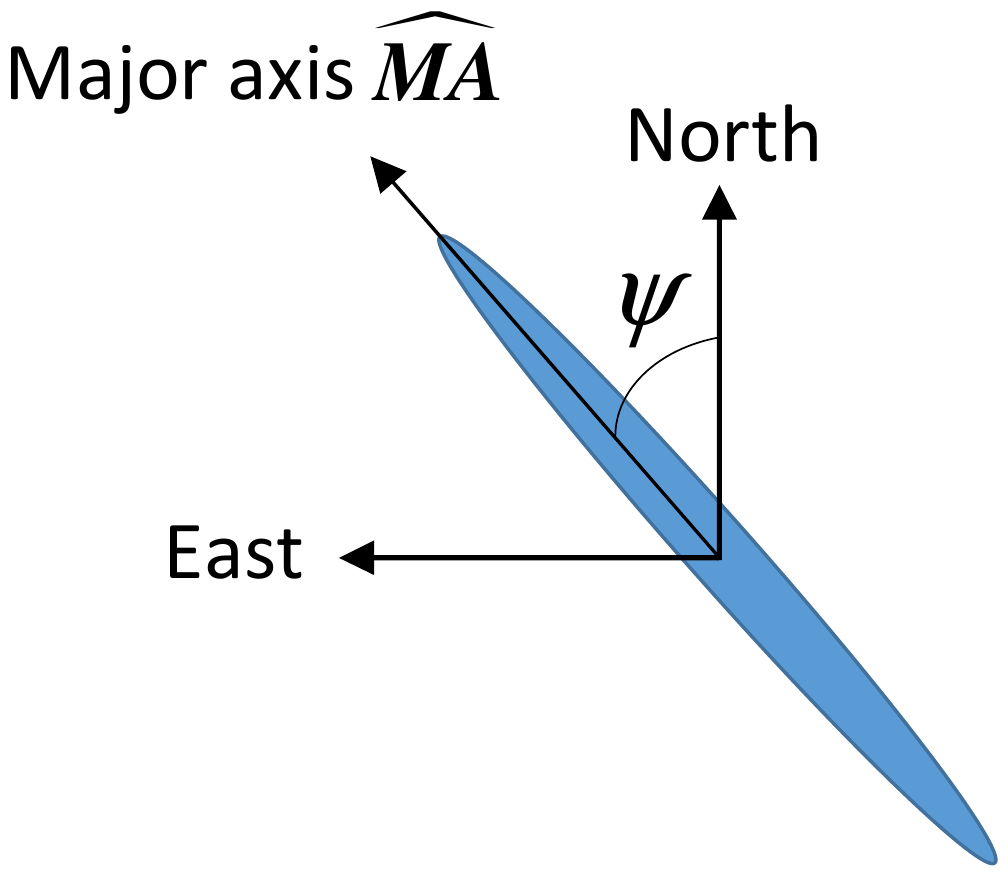}
  	\end{minipage}\hfill
  	\begin{minipage}[t]{0.28\textwidth}
		\vspace{0pt}
  		\caption{This is how an external disk galaxy like M31 appears on our sky. The direction towards the centre of its image (into screen) is $\widehat{\bm r}_{_{M31}}$. Its position angle $\psi$ is defined as the angle of its major axis $\widehat{\bm{MA}}$ eastwards of the local North (Equation \ref{North_direction}). Its inclination $i$ to the sky plane can be determined from the ellipticity of its image.}
		\label{Position_angle_diagram}
  	\end{minipage}
\end{figure}

We begin by describing how we find the angular momentum direction of the M31 disk $\widehat{\bm h}_{_{M31}}$, where we define the unit vector $\widehat{\bm v} \equiv \frac{\bm v}{\left| \bm v \right|}$ for any vector $\bm v$. Based on the ellipticity of its image, we know the inclination $i$ of the M31 disk to the plane of our sky. The orientation of this image is described by a position angle $\psi$, whose meaning is illustrated in Figure \ref{Position_angle_diagram}. Our adopted values for these parameters are given in Table \ref{M31_image_parameters}, the caption of which contains the relevant references.

The major axis of the M31 image corresponds to the direction $\widehat{\bm{MA}} \propto \widehat{\bm r}_{_{M31}} \times \widehat{\bm h}_{_{M31}}$, which is orthogonal to both the direction $\widehat{\bm r}_{_{M31}}$ towards M31 and to $\widehat{\bm h}_{_{M31}}$ as it must lie within both the sky and M31 disk planes. This leaves two possible directions for $\widehat{\bm{MA}}$. The convention is to take the one most nearly pointing east, whose local direction $\widehat{\bm E}$ and that of the local north $\widehat{\bm N}$ are
\begin{eqnarray}
	\widehat{\bm E} ~&=&~ \frac{\widehat{\bm{NCP}} \times \widehat{\bm r}_{_{M31}}}{\left| \widehat{\bm{NCP}} \times \widehat{\bm r}_{_{M31}} \right|} ~~\text{at M31 position} \\
	\widehat{\bm N} ~&=&~ \widehat{\bm r}_{_{M31}} \times \widehat{\bm E}
	\label{North_direction}
\end{eqnarray}

$\widehat{\bm E}$ is orthogonal to $\widehat{\bm r}_{_{M31}}$ and to $\widehat{\bm{NCP}}$, the direction of the North Celestial Pole. Knowing $\widehat{\bm E}$ fixes the choice of $\widehat{\bm{MA}}$ because of the convention that $\widehat{\bm{MA}} \cdot \widehat{\bm E} \geq 0$. This allows us to determine the position angle of M31
\begin{eqnarray}
\psi ~=~ \cos^{-1} \left( \widehat{\bm N} \cdot \widehat{\bm{MA}} \right) ~\text{, } ~0 \leq \psi <180^\circ
\label{Position_angle_equation}
\end{eqnarray}

The inclination of M31 is the angle of its disk normal to the line of sight towards its centre, $\widehat{\bm r}_{_{M31}}$. Thus,
\begin{eqnarray}
    i ~=~ \cos^{-1} \left|\widehat{\bm r}_{_{M31}} \cdot \widehat{\bm h}_{_{M31}} \right| ~\text{, } ~0 \leq i \leq 90^\circ
\label{Inclination_equation}
\end{eqnarray}

\begin{table}
 \begin{tabular}{lll}
	\hline
  Variable & Meaning & Value \\ 
  \hline
  $\widehat{\bm r}_{_{M31}}$ & Direction to M31 now & $\left( 121.57^\circ,-21.57^\circ \right)$ \\
  $i$ & Inclination of M31 & $77.5^\circ$ \\
   & disk to sky plane & \\
  $\psi$ & Position angle of M31 disk& $37.7^\circ \pm 0.9^\circ$ \\
   & on sky (Figure \ref{Position_angle_diagram})& \\ [3pt]
   $\widehat{\bm h}_{_{M31}}$ & Internal angular momentum& $\left( 238.65^\circ,-26.89^\circ \right)$\\
   & direction of M31 disk & \\
  \hline
 \end{tabular} 
 \caption{Observational parameters of M31 important for this work. Its sky position in Galactic co-ordinates (latitude last) is from \citet{M31_position}. We also give its inclination \citep{M31_inclination} and position angle \citep[][Section 5.2]{Chemin_2009}. Combined with radial velocity measurements, this implies its disk has a particular spin vector $\widehat{\bm h}_{_{M31}}$, which we find using Equations \ref{Position_angle_equation} and \ref{Inclination_equation}. By convention, $\widehat{\bm h}_{_{MW}}$ points towards the South Galactic Pole.}
 \label{M31_image_parameters}
\end{table}

These constraints on $i$ and $\psi$ can be satisfied if we reverse the sense in which M31 rotates (i.e. $\widehat{\bm h}_{_{M31}} \to -\widehat{\bm h}_{_{M31}}$). Its actual sense of rotation must be determined observationally. In Galactic co-ordinates, the northern part of M31 is receding from us relative to its southern part, indicating that its angular momentum must point further east. Thus, the Galactic longitude of $\widehat{\bm h}_{_{M31}}$ must exceed that of M31 itself (by $<180^\circ$). Combined with the other constraints, this unambiguously determines $\widehat{\bm h}_{_{M31}}$.

We used the 2D Newton-Raphson algorithm to vary the Galactic latitude and longitude of $\widehat{\bm h}_{_{M31}}$ in an attempt to match the available observational constraints on $i$ and $\psi$ (Table \ref{M31_image_parameters}). Starting from a guess in the correct hemisphere, our algorithm converged on the same solution as that in Table 4 of \citet{Raychaudhury_1989}, providing an important cross-check. However, no explanation was given there for how $\widehat{\bm h}_{_{M31}}$ was derived or the assumed M31 position angle and disk inclination, both of which we use more recent measurements for.

\subsection{The MW-M31 Orbital Plane}
\label{Orbital_plane_finding}

The satellites of the MW mostly lie within a thin plane and co-rotate within it \citep{Kroupa_2013}. The same is true for M31 \citep{Ibata_2013}.\footnote{Co-rotation can only be proved with proper motions, but it is strongly suggested by a radial velocity gradient.} We investigate the scenario where these satellite planes were formed by a past close encounter between the MW and M31. Such an encounter is inevitable in MOND \citep{Zhao_2013} but impossible in $\Lambda$CDM as dynamical friction between their dark matter halos would cause a rapid subsequent merger \citep[e.g.][]{Privon_2013}. This difference between the theories may provide a basis for distinguishing between them \citep{Kroupa_2015}.

We use a simple toy model to constrain the MW-M31 orbital angular momentum direction $\widehat{\bm h}$ required by this scenario. Our model is based on two simplifying assumptions $-$ the tidal torque exerted by M31 on the MW is assumed to act only at the time of their closest approach and only on the part of the MW closest to M31 at that time (and vice versa). We also assume that $\widehat{\bm r}_{_{M31}}$ has rotated by an angle $\phi \approx 125^\circ$ since that time \citep[][Figure 9]{Belokurov_2014}. Our calculations suggest that the actual value is very likely within $6^\circ$ of this.

By definition, the present direction towards M31 must be orthogonal to $\widehat{\bm h}$, constraining $\widehat{\bm h}$ to lie along a great circle. We measure position along this great circle using the angle $\theta$ measured southwards from the point on it in the northern Galactic hemisphere at a Galactic longitude of $180^\circ$.

In our model, the tidal torque exerted on galaxy $i$
\begin{eqnarray}
	\Delta \bm h_i \propto \left( \widehat{\bm h}_i \times \widehat{\bm r} \right)\left( \widehat{\bm h}_i \cdot \widehat{\bm r} \right) ~~\text{where}~ i = \text{MW or M31}
	\label{Tidal_torque_direction}
\end{eqnarray}

Galaxy $i$ has its disk angular momentum in the direction $\widehat{\bm h}_i$ while $\widehat{\bm r}$ is the direction towards the other galaxy \emph{at the time of their closest approach}. Equation \ref{Tidal_torque_direction} shows that we do not expect there to be much tidal torque on a galaxy if $\widehat{\bm r}$ is either within its disk plane or along its disk normal. Although this is not totally accurate, it does suggest that such solutions would have difficulty in explaining the large amount of tidal torque required to create the satellite plane of at least one major LG galaxy given the significant observed disk-satellite plane misalignment for both the MW and M31 (Figure \ref{Directions_plotting_3D}). Thus, we only consider solutions satisfying
\begin{eqnarray}
	\cos 87^\circ < \left| \widehat{\bm h}_i \cdot \widehat{\bm r} \right| < \cos 3^\circ ~~\text{for both}~ i = \text{MW and M31}
	\label{h_limits}
\end{eqnarray}

The material which eventually forms the satellite plane around galaxy $i$ has an angular momentum parallel to
\begin{eqnarray}
	\widehat{\bm h}_i ~+~ \left( \tan \kappa_{_{i}} \right) \widehat{\Delta \bm h}_i
\end{eqnarray}

\begin{figure}
	\centering 
	\includegraphics [width = 5.5cm] {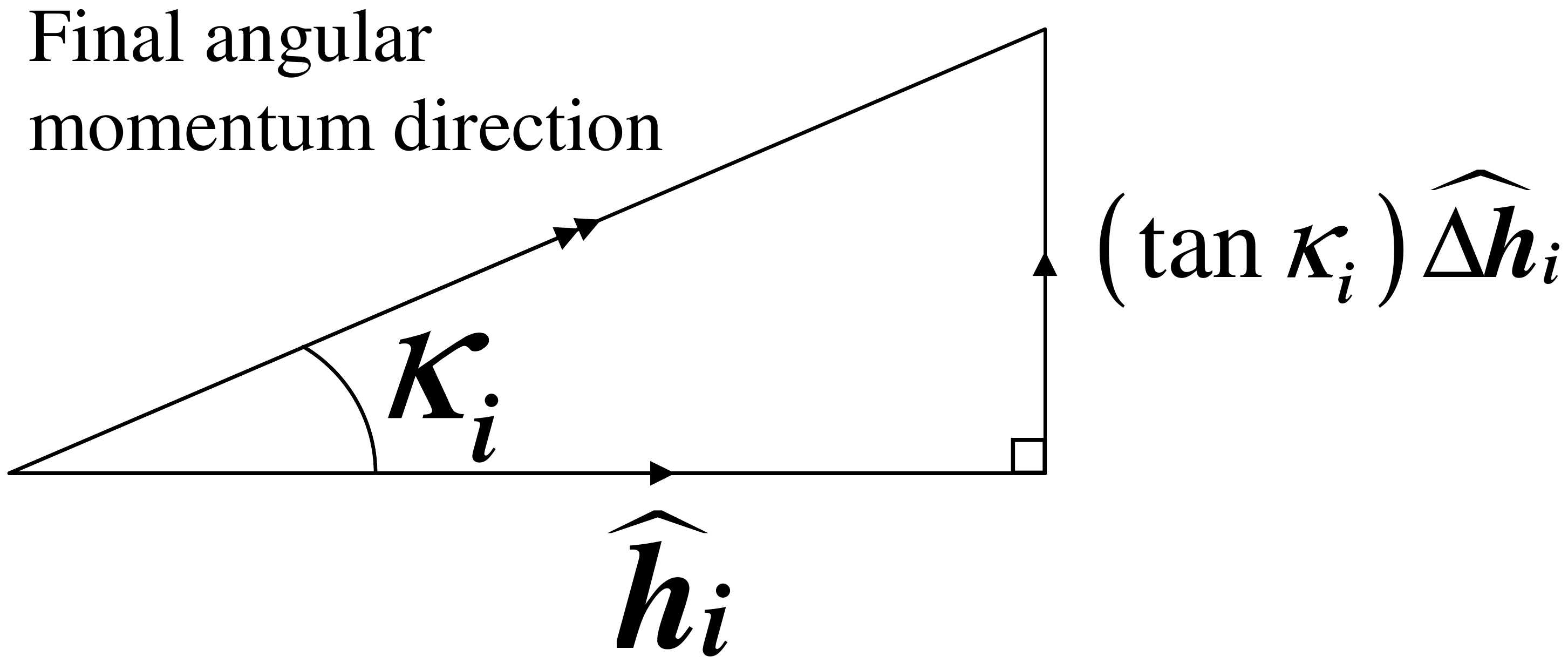}
	\caption{Illustration of how the MW-M31 interaction affects the angular momentum of material in the outer disk of galaxy $i$. In our model, the tidal torque on it is orthogonal to its original angular momentum $\widehat{\bm h}_i$ arising from disk rotation. As a result, the disk-satellite plane mismatch angle $\kappa_{_i}$ measures the ratio of angular momentum gained to that originally present.}
	\label{Tidal_torque_diagram}
\end{figure}

The parameter $\tan \kappa_{_{i}}$ governs the relative importance of the tidal torque on galaxy $i$ and the angular momentum its spinning disk already possessed before the interaction. Because the unit vectors $\widehat{\Delta \bm h}_i$ and $\widehat{\bm h}_i$ are orthogonal, $\tan \kappa_{_{i}}$ determines the model-predicted angle $\kappa_{_{i}}$ between the orientations of the disk and dominant satellite plane of galaxy $i$ (Figure \ref{Tidal_torque_diagram}). Without a more detailed model, it is difficult to estimate this angle. We assume that its tangent is in the range ${\left(0.1-10\right)}$ and allow it to be different for the MW and M31 due to their different masses, disk sizes and rotation speeds.


\begin{figure}
	\centering 
		\includegraphics [width = 8.5cm] {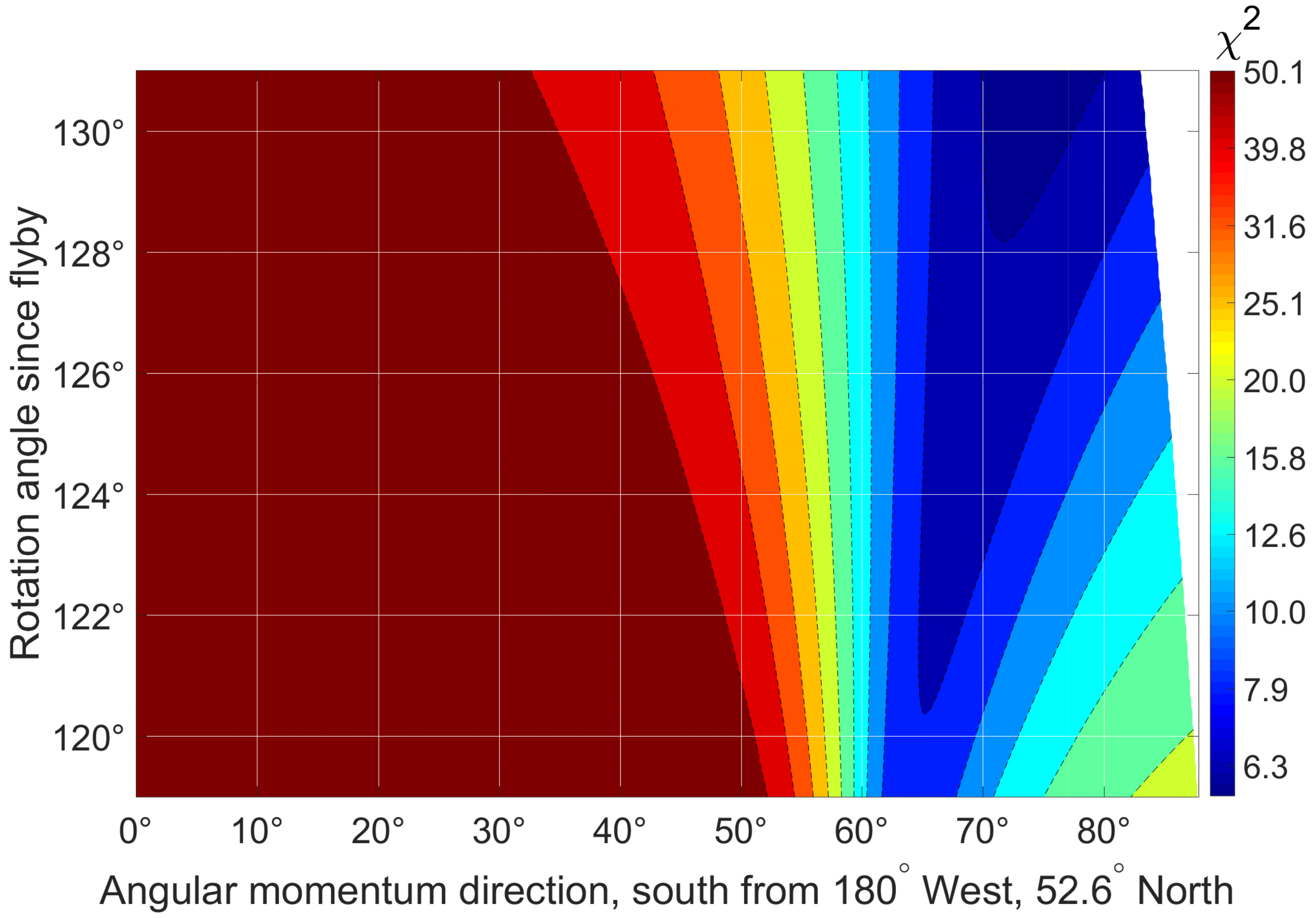}
		\includegraphics [width = 8.5cm] {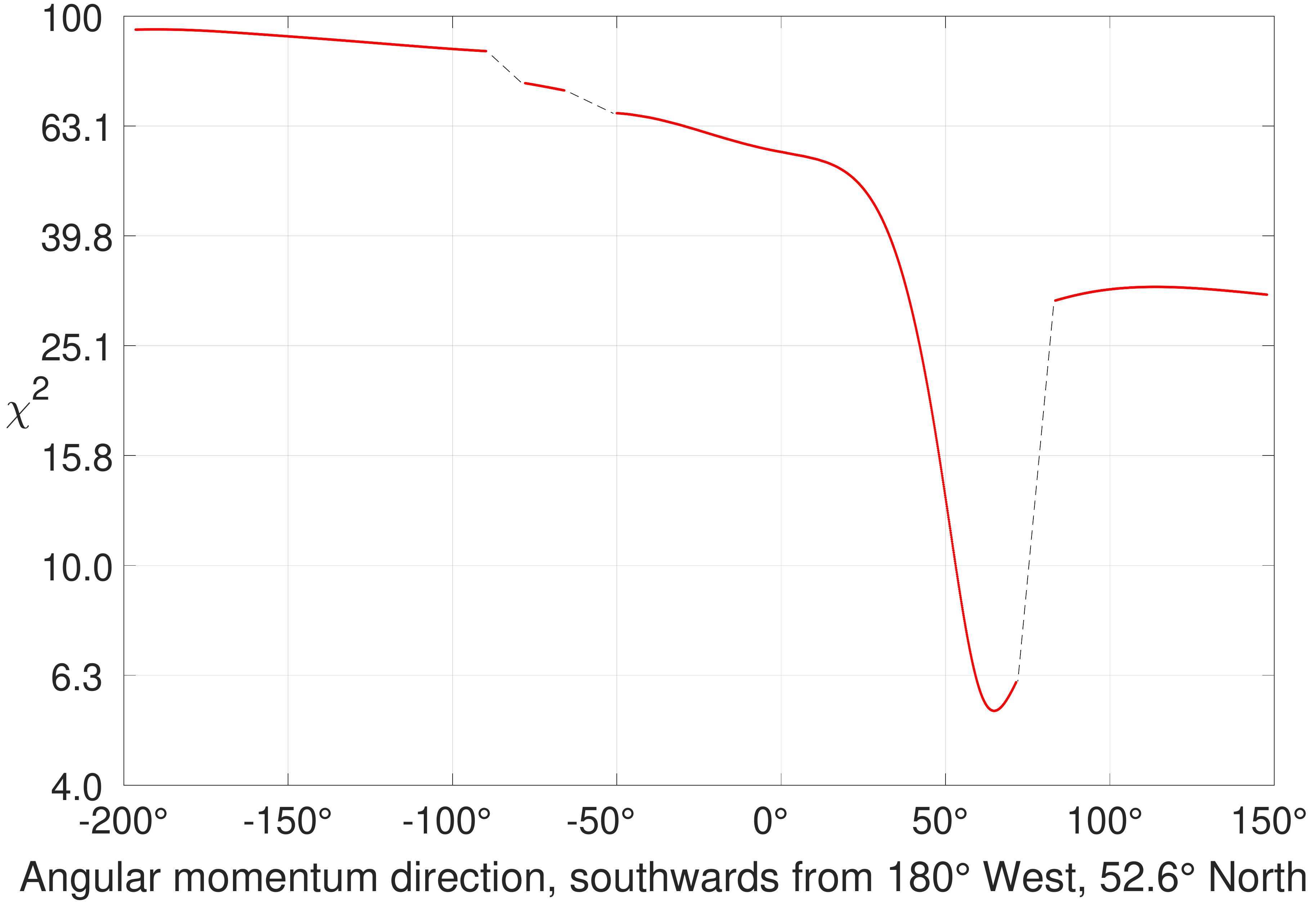}
		\caption{\emph{Top:} Goodness of fit $\chi^2$ of our toy model to the MW and M31 satellite plane orientations as a function of model parameters, assuming an uncertainty of $10^\circ$ for both systems. The best-fitting solution is given in Table \ref{Directions_table}. The gap arises as some models violate Equation \ref{h_limits} and so could not plausibly lead to enough tidal torque on at least one major LG galaxy to explain the misalignment between its disk and dominant satellite plane (Figure \ref{Directions_plotting_3D}). The issue of most relevance is that the MW seems to have been almost within the M31 disk plane at the time of their closest approach. \emph{Bottom}: $\chi^2$ as a function of the adopted MW-M31 orbital plane with a fixed rotation angle of $\phi = 131^\circ$ since their flyby. The dashed black lines correspond to directions of $\widehat {\bm h}$ for which it is not possible to satisfy the 4 constraints imposed by Equation \ref{h_limits}, leading to 4 distinct excluded ranges of $\theta$ (one of which has been folded around the figure for clarity).}
\label{Flyby_geometry_grid}
\end{figure}

For every value of $\theta$, we vary the rotation angle $\phi$ between $119^\circ$ and $131^\circ$. Each time, we find the value of $\tan \kappa_{_{MW}}$ that minimises the angle between the calculated and observed spin vector of the MW satellite plane. The same procedure is used for M31. We combine these angular differences in quadrature to obtain a $\chi^2$ statistic, which we base on an allowance of $10^\circ$ for both satellite planes. Our results are shown in Figure \ref{Flyby_geometry_grid}.

Restricting to angles $\phi \leq 131^\circ$, we can obtain a solution with $\chi^2 = 5.0$ for $\theta = 75^\circ$, $\phi = 131^\circ$. Although the resulting $\chi^2$ is a little higher than 2, it is still quite acceptable. Our toy model is thus able to provide a plausible explanation for the origins of the MW and M31 satellite planes based on these galaxies having undergone a past close encounter. Naturally, we hope to refine our model in future, perhaps by using the RAyMOND algorithm \citep{Candlish_2015} or the publicly available Phantom of RAMSES algorithm \citep{PoR}, both of which handle MOND explicitly by adapting the RAMSES algorithm \citep{Teyssier_2002}. It is already possible to use the latter to simulate interacting disk galaxies in MOND \citep{Thies_2016}.

We find that it is much more difficult to explain the orientation of the M31 satellite system than that of the MW. This may be related to the M31 satellite plane being inclined to its disk by ${\ssim 47^\circ}$ \citep{Ibata_2013} whereas the MW satellite plane is almost polar with respect to its disk \citep{Kroupa_2013}. This makes it more likely that the M31 satellite plane has precessed from its initial orientation \citep{Fernando_2016}, especially as the M31 disk has a scale length ${\ssim 2.5 \times}$ larger than that of the MW \citep{Courteau_2011, Bovy_2013}. However, such precession effects would tend to thicken the M31 satellite plane as they would be weaker for satellites further from M31 \citep{Fernando_2016}. The very low observed thickness \citep[${12.6 \pm 0.6}$ kpc,][]{Ibata_2013} therefore argues against such an explanation.

Due to the different disk-satellite plane misalignments for the MW and M31, our model implies that the MW was more affected by tides from M31 than vice versa (Table \ref{Directions_table}). In MOND, we can obtain a good estimate of the mass of a galaxy from its rotation speed $v_{_f}$ in the flat outer region of its rotation curve (Equation \ref{BTFR}). This suggests that the slower-rotating MW with $v_{_f} = 180$ km/s \citep{Kafle_2012} has a lower mass than the faster-rotating M31 with $v_{_f} = 225$ km/s \citep{Carignan_2006}. As the MW disk is also easier to tilt because it has less specific angular momentum than the M31 disk, it is reasonable to get ${\tan \kappa_{_{MW}}}$ a few times larger than ${\tan \kappa_{_{M31}}}$. Therefore, in MOND, the faster asymptotic rotation speed of M31 is directly related to the larger observed angle between the MW disk and its satellite plane compared to the same quantity for M31 (Figure \ref{Directions_plotting_3D}).\footnote{These observational facts are otherwise unrelated.} This is exacerbated by our model implying that $\widehat{\bm r}_{_{M31}}$ at the time of the flyby lay only ${\ssim 6^\circ}$ out of the M31 disk plane, reducing how much torque could be exerted on it (Equation \ref{Tidal_torque_direction}). At that time, $\widehat{\bm r}_{_{M31}}$ was ${\ssim 66^\circ}$ from the MW disk plane. However, the larger scale length of the M31 disk may have counteracted these factors somewhat. Combining these considerations suggests that
\begin{eqnarray}
\label{k_ratio_estimate}
	\frac{\tan \kappa_{_{MW}}}{\tan \kappa_{_{M31}}} &\approx& \left( \frac{v_{_{f, MW}}}{v_{_{f,M31}}} \right)^5 \frac{r_{_{d,MW}}}{r_{_{d,M31}}} \frac{\sin 66^\circ \cos 66^\circ}{\sin 6^\circ \cos 6^\circ} \\
	&\approx& 4
\end{eqnarray}

The ratio between the disk scale lengths $r_{_d}$ of the MW and M31 may have been different in the past and their orientations may have been slightly different too. Moreover, our model is only a very basic one. Despite this, Equation \ref{k_ratio_estimate} suggests that $\tan \kappa_{_{MW}} \approx 4 \tan \kappa_{_{M31}}$, similar to the ratio in our best-fitting model (Table \ref{Directions_table}).


\begin{table}
 \begin{tabular}{ll}
\hline
Quantity & Value \\ \hline
  MW satellite plane spin vector & $\left( 176.4^\circ, -15.0^\circ \right)$\\
  M31 satellite plane spin vector & $\left( 206.2^\circ, 7.8^\circ \right)$\\ [5pt]
  MW-M31 orbital angular & $75^\circ$\\
  momentum direction $\theta$ & \\
  Expected MW-M31 orbital pole $\widehat {\bm h}$ & $\left( 217.3^\circ, -15.1^\circ \right)$ \\
Rotation angle of MW-M31 & $131^\circ$\\
line since their flyby, $\phi$ & \\ [5pt]
$\tan \kappa_{_{MW}}$ (see Figure \ref{Tidal_torque_diagram}) & 3.7\\
$\tan \kappa_{_{M31}}$ (see Figure \ref{Tidal_torque_diagram}) & 1.0\\ [5pt]
  \hline
 \end{tabular}
 \caption{Values of the quantities most relevant to the geometry of a past MW-M31 flyby and its effects. The MW satellite plane orientation is from \citet[][Section 3]{Kroupa_2013} while that of M31 is from their Section 4. Other parameters are determined using a toy model of a past MW-M31 interaction forming their planes of satellite galaxies (Section \ref{Orbital_plane_finding}).}
 \label{Directions_table}
\end{table}

For both major LG galaxies, the rather large values of $\kappa$ suggest that some material may have been pulled out of them and become unbound. This may explain why some LG non-satellite galaxies have unusual kinematics \citep{Banik_Zhao_2017}. It is possible that the material in some of them was once part of the MW or M31 disk or in their satellite system. Further study of this scenario must be left for future works.

\section{The Local Group in MOND}
\label{MOND_simulation}

\subsection{Governing equations}
\label{MOND_governing_equations}

We now conduct a more detailed simulation of the LG in MOND, taking advantage of the MW-M31 orbital pole determined in Section \ref{Orbital_plane_finding}. The MW-M31 trajectory is simulated by advancing them according to their mutual gravity supplemented by the cosmological acceleration term \citep[e.g.][Equation 24]{Banik_Zhao_2016}.
\begin{eqnarray}
      \label{MW_M31_governing_equation}
	\ddot{\bm{r}}_{_{rel}} &=& \bm{g_{_{M31}}} - \bm{g_{_{MW}}} + \frac{\ddot{a}}{a}\bm{r}_{_{rel}} ~\text{  where} \\
	\bm{r_{_{rel}}} &\equiv & \bm{r_{_{M31}}} - \bm{r_{_{MW}}}
\end{eqnarray}

Here, the cosmic scale-factor is $a \left(t\right)$ and $\bm{r_{_i}}$ is the position vector of galaxy $i$ (MW or M31), at whose location the gravitational field (excluding self-gravity) is $\bm{g_{_i}}$. We use an overdot to indicate the time derivative of any quantity $q$ e.g. ${\dot {q} \equiv \frac{\partial q}{\partial t}}$. All position vectors are with respect to the LG barycentre, which we take to be 0.3 of the way from M31 towards the MW. This is based on the asymptotic rotation curve of the MW flatlining at ${\ssim 180}$ km/s \citep{Kafle_2012} while the equivalent value for M31 is ${\ssim 225}$ km/s \citep{Carignan_2006}. In the context of MOND, this suggests that the mass of M31 is $\left( \frac{225}{180} \right)^4 \approx {2.3 \times}$ that of the MW (Equation \ref{BTFR}).

Although MOND is known to work well at explaining the internal dynamics of galaxies outside the LG \citep[e.g.][]{Famaey_McGaugh_2012}, we should check if this is the case for the MW and M31 before using it to determine the gravity they exert on each other and on the rest of the LG. For our neighbour M31, MOND can provide a fairly good match to its rotation curve using its observed baryonic distribution \citep[][Figure 4]{Corbelli_2007}. For our work, it is important to note that this fit remains good out to rather large radii (${\ssim 35}$ kpc or 7 disk scale lengths). A similar analysis for the MW is complicated slightly by our position within its disk. However, it has recently become clear that MOND can explain its rotation curve fairly well \citep{McGaugh_2016_MW} and even provides a good match to its escape velocity curve \citep{Banik_2017_escape}. Thus, applying Equation \ref{BTFR} to the MW and M31 rotation curves yields reasonable estimates for their baryonic masses, the vast majority of which resides in stars \citep[91\% for M31 and 81\% for the MW,][Section 2.2]{Yin_2009}. This is almost certainly not representative of the Universe as a whole given that most of the mass in galaxy clusters is hot gas that has only recently been discovered at X-ray wavelengths \citep[e.g.][]{Vikhlinin_2006}. Indeed, the location of all the baryons in the Universe is far from certain, with significant amounts perhaps residing in an even more diffuse form \citep{Nicastro_2008}.

Having obtained MW and M31 masses ($M_{MW}$ and $M_{M31}$) in this way, we treat them as point masses and find the gravitational field $\bm{g}$ they exert at position $\bm{r}$ using the quasilinear formulation of MOND \citep{QUMOND}. We assume the `simple' interpolating function between the Newtonian and deep-MOND regimes \citep{Famaey_Binney_2005} that works best for the MW rotation curve \citep{Iocco_Bertone_2015}.
\begin{eqnarray}
	\bm{g_{_N}} &\equiv & -\sum_{i=MW,M31} \frac{GM_{i} \left( \bm{r} - \bm{r_{_{i}}} \right)}{ \left|\bm{r} - \bm{r_{_{i}}} \right|^3}  \\
	\nabla \cdot \bm{g} &\equiv & \nabla \cdot \left[\nu \left( \frac{\left| \bm{g_{_N}} \right|}{a_{_0}}\right) \bm{g_{_N}} \right] ~~\text{ where} \\
	\nu \left( x \right) &=& \frac{1}{2} ~+~ \sqrt{\frac{1}{4} + \frac{1}{x}}
	\label{Simple_interpolating_function}
\end{eqnarray}

The appropriate boundary conditions are similar to Newtonian gravity, but for definiteness we give them here.
\begin{eqnarray}
	\nabla \times \bm{g} &=& 0 \\
	\bm{g} &\to & 0 ~~\text{as } \left| \bm{r} \right| \to \infty
\end{eqnarray}

We use direct summation to obtain $\bm{g}$ from its divergence.
\begin{eqnarray}
\bm{g \left( \bm{r} \right)} ~=~ \int \nabla \cdot \bm{g} \left( \bm{r'}\right) \frac{\left( \bm{r} - \bm{r'} \right)}{|\bm{r} - \bm{r'}|^3} ~d^3\bm{r'}
\end{eqnarray}

$\nabla \cdot \bm{g}$ is calculated out to almost ${150 \times}$ the MW-M31 separation, beyond which it should be very nearly spherically symmetric. Due to the shell theorem, it is unnecessary to consider $\nabla \cdot \bm{g}$ (or `phantom dark matter') at even larger radii. As we only determine $\bm{g}$ out to ${66.5 \times}$ the MW-M31 separation, our results should be nearly free of edge effects. At larger distances, we assume that the MW and M31 can be treated as a single point mass located at their barycentre, yielding $\bm{g} = \nu \bm{g_{_N}}$.

\subsection{MW-M31 trajectory}

Using the gravitational field thus found, we integrate the MW-M31 trajectory backwards from present conditions. In general, the galaxies will not be on the Hubble flow at the start time of our simulations $t_{_i}$, when the cosmic scale-factor $a_{_i} = 0.05$ and the Hubble parameter $H \equiv \frac{\dot{a}}{a}$ is $H_{_i}$. However, deviations from the Hubble flow are observed to be very small at early times \citep{Planck_2013}. In order to satisfy this condition at $t_{i}$, we vary the total mass of the MW and M31 using a Newton-Raphson root-finding algorithm. This ensures that
\begin{eqnarray}
	\dot{\bm{r}}_{_{rel}} = H_{_i} \bm{r_{_{rel}}} ~\text{ when } t = t_{_i}
      \label{Hubble_flow_initial}
\end{eqnarray}

The MW and M31 are not on a purely radial orbit. Their mutual orbital angular momentum prevents them from converging onto the Hubble flow at very early times. This is unrealistic as any non-radial motion must have arisen due to tidal\footnote{affecting the MW and M31 differently} torques well after the Big Bang. Thus, we take the MW-M31 orbit to be purely radial prior to their first turnaround at $t \approx 3$ Gyr. After this time, we assume their trajectory conserves angular momentum at its present value. This implies the MW-M31 angular momentum was gained near the time of their first turnaround, when their large separation would have strengthened tidal torques. At later times, the larger scale factor would weaken tidal torques, suggesting that these have a much smaller effect around the time of the second MW-M31 turnaround than the first.

At present, there is no detailed theory for structure formation in MOND because it is unclear how to apply it to regions only slightly denser than the cosmic mean density. Assuming a particular model, structure formation was found to be more efficient than in $\Lambda$CDM \citep{Llinares_2008}. Observationally, there are several indications that this is actually the case \citep{Peebles_2010}, including the rather high fraction of pure (bulgeless) disk galaxies \citep{Kormendy_2010}.

Thus, structure formation in MOND $-$ and perhaps in the Universe $-$ is not so reliant on growth at late times through mergers, which would be less efficient in MOND due to the absence of dynamical friction between extended dark matter halos. Instead, galaxies would form relatively rapidly after the Big Bang, emptying their surroundings due to the strong long-range gravitational attraction in the model. This would lead to many widely separated `island universes' with fairly empty intervening voids, perhaps similar to the Local Volume (out to 8 Mpc) which does seem to contain voids emptier than might be expected in $\Lambda$CDM \citep{Tikhonov_2009}. With mass draining on to a few well-separated massive galaxies, it would be natural for the MW and M31 to end up fairly isolated. Thus, there would not be much tidal torque on the MW-M31 system, leaving its orbit close to radial. This suggests that it would not be all that unusual for us to find ourselves in a galaxy which had a past close encounter with its nearest large neighbour if structure formation proceeded more efficiently than in $\Lambda$CDM. Of course, it remains to be seen whether it forms too efficiently in MOND.


Given the way we expect structure to form in MOND, we assume the MW and M31 masses do not grow by accretion at late times. However, an important effect included in our models is a 5\% reduction in their masses at the time of closest approach, when their simulated separation was just 14.2 kpc.\footnote{This depends on the present proper motion of M31, for which only an upper limit is available \citep{M31_motion}. Thus, the closest approach distance is unlikely to exceed 50 kpc but can be made arbitrarily small \citep{Zhao_2013}.} Considering that the MW disk has a scale length of 2.15 kpc \citep{Bovy_2013} while the corresponding quantity for M31 is 5.3 kpc \citep{Courteau_2011}, it is very likely that some of the mass in these galaxies would be expelled to large distances and escape from them.

\begin{figure}
	\centering 
		\includegraphics [width = 8.5cm] {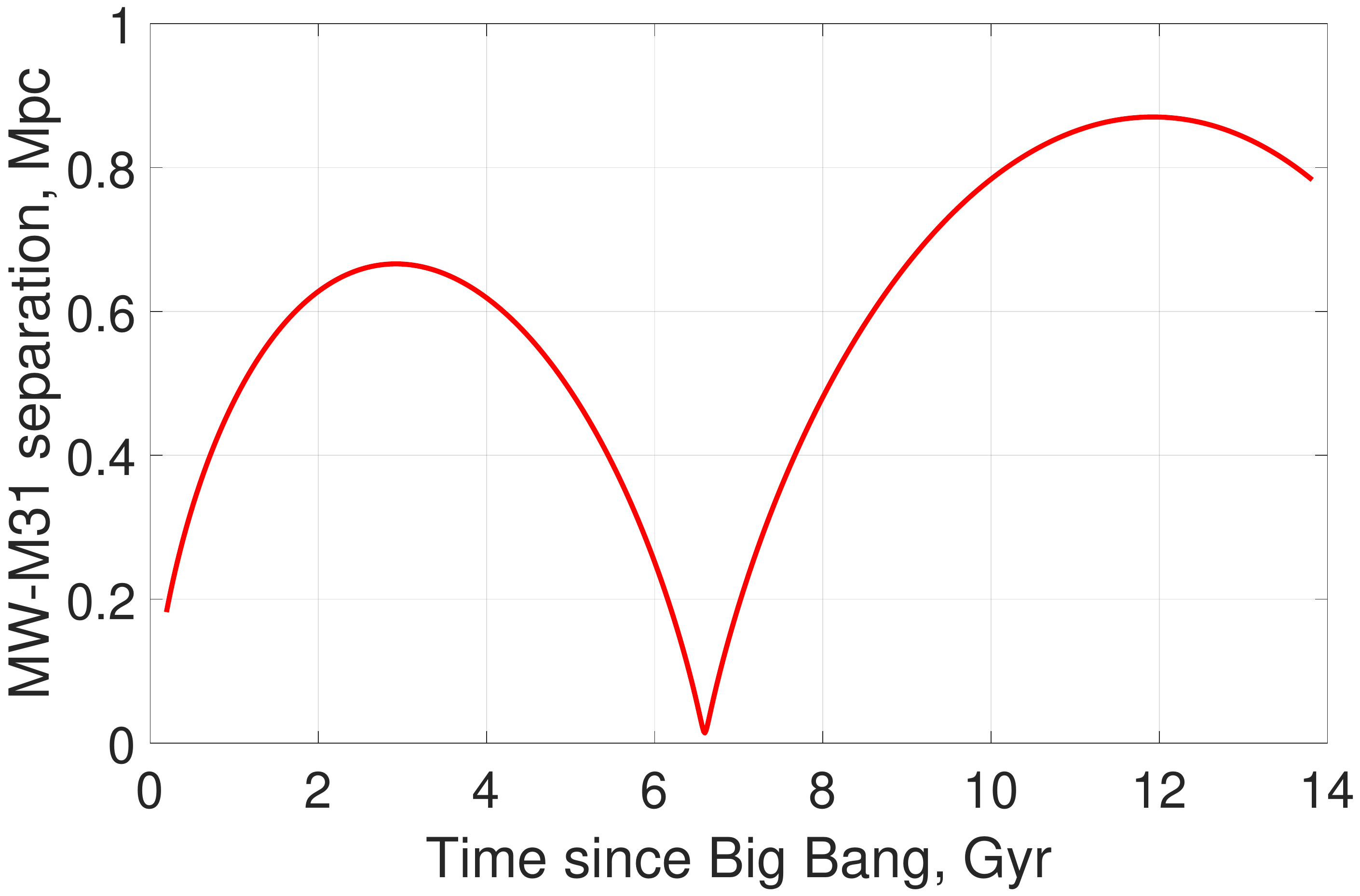}
		\caption{MW-M31 separation in our MOND simulation, showing a past close flyby 6.59 Gyr after the Big Bang at a closest approach distance of 14.17 kpc. At that time, their relative velocity was 716 km/s, of which 501 km/s was due to motion of the MW. The higher second apogalacticon is partly due to the effect of cosmology (Equation \ref{MW_M31_governing_equation}) and our assumption that the MW and M31 lose 5\% of their mass around the time of their encounter (see text).}
	\label{MW_M31_trajectory_FL}
\end{figure}

This mass could reside in the halos of hot gas surrounding each galaxy. Such a halo has been detected around M31 based on absorption features in spectra of background quasars \citep{Lehner_2015}. A similar halo is thought to be necessary around the MW to explain the truncation of the Large Magellanic Cloud's gas disk \citep{Salem_2015}. These gas halos seem to contain perhaps $3 \times 10^{10} M_\odot$ each, with much larger amounts being very unlikely given constraints from the MW escape velocity curve \citep{Banik_2017_escape}. Considering that the MW rotation curve flatlines at $\ssim 180$ km/s \citep{Kafle_2012} while that of M31 flattens at $\ssim 225$ km/s \citep{Carignan_2006}, MOND suggests their total baryonic mass is $2.3 \times 10^{11} M_\odot$. This makes it quite feasible for them to have lost ${10^{10}} M_\odot$ of hot gas around the time of their encounter, as our model implies. Some hot gas in an extended halo could also explain why the rotation curve-based estimate of the total MW and M31 mass falls a little below our timing argument estimate of $2.9 \times 10^{11} M_\odot$ (for simplicity, we fix the MW:M31 mass ratio at 3:7 and scale up their masses slightly to make the timing argument work).

There are several other aspects of the problem which we include in our model using techniques we developed. We defer a more detailed explanation of our procedures to a forthcoming publication which will investigate the MW-M31 trajectory in MOND. For the present contribution, the major result is that Equation \ref{Hubble_flow_initial} can be satisfied by backwards integration from present conditions using MW and M31 masses consistent with their rotation curves in MOND (themselves consistent with observed baryonic disk masses) and the more extended halos of hot gas that have recently been detected around them \citep{Lehner_2015, Nicastro_2016}. The resulting MW-M31 trajectory is shown in Figure \ref{MW_M31_trajectory_FL}. A past close encounter is inevitable in the context of MOND \citep{Zhao_2013} due to their slow relative tangential motion \citep{M31_motion} and the strong gravity in this model. We previously discussed how the thick disk of the MW and the LG satellite planes may well have formed due to this interaction (Sections \ref{Introduction} and \ref{Orbital_plane_finding}, respectively). Here, we consider its effect on the rest of the LG.


\subsection{Test particle trajectories}

Once the MW-M31 trajectory is known, we can determine the gravitational field $\bm{g}$ everywhere within the LG at all times under the assumption that only these point masses are present in an otherwise homogeneous Universe. This allows us to advance the trajectories of test particles according to
\begin{eqnarray}
	\label{Test_particle_acceleration}
	\ddot{\bm{r}} ~&=&~ \frac{\ddot{a}}{a}\bm{r} ~+~ \bm{g} \\
	\dot{\bm{r}} ~&=&~ H_{_i}r ~\text{ when }t = t_{_i}	
\end{eqnarray}

Although the particles could not have started exactly on the Hubble flow, we consider this a reasonable assumption for reasons we now discuss. The MW and M31 could not have formed much earlier than when $a = 0.05$ because the MOND free-fall collapse time on to a point mass is given by
\begin{eqnarray}
	t_{_{ff}} &=& \frac{r_{_0}}{v_{_f}} \sqrt{\frac{\pi}{2}}
\end{eqnarray}

$v_{_f}$ is given by Equation \ref{BTFR} and taken to be 180 km/s for the MW \citep{Kafle_2012}. Assuming the material currently in it must have turned around from a distance $\ga 100$ kpc (Equation \ref{r_exc}), this yields a free-fall timescale of $t_{_{ff}} = {540}$ Myr. This is much more than the 191 Myr age of the Universe when $a = 0.05$ in standard cosmology, suggesting that it is not appropriate to start our simulations much earlier.

Moreover, even if the MW existed as a point mass since the Big Bang, the resulting peculiar velocity some co-moving distance $d$ away would be
\begin{eqnarray}
	\label{v_pec}
	av_{pec} ~\equiv ~ a \overbrace{\left( \dot{r} - Hr \right)}^{v_{pec}} &=& \int_0^t a g~dt ~~\text{where} \\
	g &=& \frac{\sqrt{GMa_{_0}}}{ad}
\end{eqnarray}

The integrating factor $a \left( t \right)$ accounts for the effects of Hubble drag. We have assumed that the particle nearly follows the Hubble flow so that its distance to the mass can be taken as $a\left( t \right) d$. At a co-moving distance of 2 Mpc, the peculiar velocity gained would only be 65 km/s by the time $a = 0.05$. At that time, the Hubble velocity of the particle was nearly 340 km/s, justifying our assumption that it was almost on the Hubble flow.

To understand the effect on the present-day velocity field, we must bear in mind that both the present position and velocity of a test particle would be affected if we get its velocity wrong at some earlier time. Thus, if we wish to know the velocity at a particular position today, we would need to consider a different test particle starting at a different position. This `initial condition drag' scales down the effect of a velocity error at earlier times (when the scale factor was $a$) on the present velocity \emph{at fixed position} by a factor of ${\sim a^{2.4}}$ \citep[][Figure 4]{Banik_Zhao_2016}. However, even if we make the more conservative assumption that it simply scales with $a$ (like traditional Hubble drag), a 65 km/s velocity error when $a = 0.05$ would only affect the present LG velocity field by ${\ssim 3}$ km/s. This is probably why our axisymmetric dynamical model of the LG was hardly affected by using a different start time \citep[][Section 4.6]{Banik_Zhao_2016}. Thus, our choice of initial conditions should be sufficient to get an approximate idea of how the LG might have been affected by a past MW-M31 flyby. Moreover, the fact that motions at high redshift have only a weak impact on our results implies that they should be robust to uncertainties surrounding the application of MOND in a cosmological context (at lower redshifts, the MW and M31 have already formed and so better approximate the isolated situations where it is clear how MOND works).


Because the MW and M31 must have accreted matter from some region prior to the start of our simulation, we exclude all test particles starting within a distance $r_{_{exc,i}}$ of galaxy $i$. We determine this by requiring that the excluded volume has as much baryonic matter as galaxy $i$, taking the density of baryons to be the cosmic mean value. This is obtained from the fraction $\Omega_{b,0} = 0.049$ that baryons currently comprise of the cosmic critical density, which we found by taking $H_{_0} \equiv H \left(t_{_0} \right) = $ 67.3 km/s/Mpc \citep[][Table 4]{Planck_2015}. The cosmic baryon density can be estimated using Big Bang nucleosynthesis arguments $-$ only a narrow range of values is consistent with the primordial abundances of light elements such as deuterium \citep{Cyburt_2016}.
\begin{eqnarray}
	\frac{4 \pi}{3}{r_{{exc,i}}}^3 \times \overbrace{\frac{3{H_{_0}}^2}{8\pi G} \Omega_{_{b,0}} {a_{_i}}^{-3}}^{\text{Baryon density at }t_{_i}} ~\equiv ~ M_i ~~ \left( \text{for } r_{{exc,i}} \right)
	\label{r_exc}
\end{eqnarray}

For consistency, it is necessary that the sizes of the excluded regions satisfy
\begin{eqnarray}
	r_{_{exc,MW}} ~+~ r_{{exc,M31}} ~\leq ~\left|\bm{r_{_{rel}}}\right| ~\text{ when } t = t_{_i}
\end{eqnarray}

This inequality applies because $r_{_{exc}}$ is 77.7 kpc for the MW and 102.5 kpc for M31, leading to a total of 180.2 kpc $-$ interestingly, this is just smaller than $r_{_{rel}} \left( t_{_i} \right) = 182.1$ kpc, suggesting that the two galaxies accreted matter from regions which just touched. This remains the case if we use a slightly different start time as ${r_{{exc,i}}} \propto a_{_i}$, similarly to $r_{_{rel}} \left( t_{_i} \right) - $ at such early times, both galaxies would follow the Hubble flow rather closely. However, this coincidence does not occur in $\Lambda$CDM, a model in which the excluded regions would very likely overlap \citep[][Section 2.2.1]{Banik_Zhao_2016}.


\subsection{Simulation results}

\begin{figure}
	\centering 
		\includegraphics [width = 8.5cm] {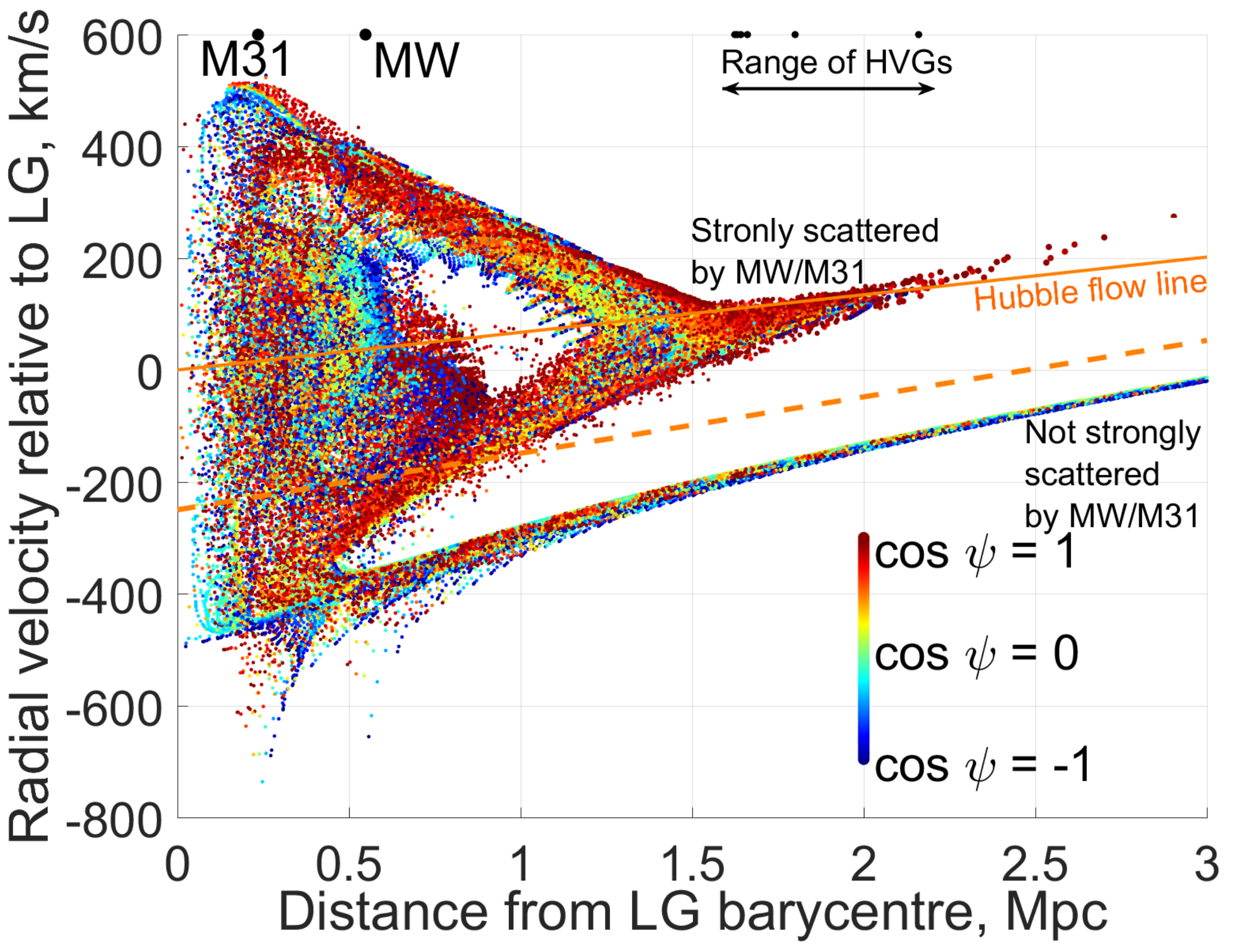}
		\caption{Hubble diagram of the test particles in our simulation coloured by their value of $\cos \psi$, which parametrises how well their orbital angular momenta align with that of the MW-M31 orbit (Equation \ref{cos_psi_definition}). We show the Hubble flow line (solid orange) and a ${1.5 \times}$ steeper line (dashed orange) which we use to select analogues of HVGs in later figures. Particles below this line have generally never interacted closely with the MW or M31, unlike particles above the line. The black dots along the top edge of the figure indicate distances to the MW, M31 and the HVGs. Marker sizes have been enlarged in some regions for clarity, but do not correspond to volume factors (see text). Thus, particles at distances ${\la 1}$ Mpc represent much less mass than it might appear.}
	\label{Hubble_diagram_coloured_FL}
\end{figure}

\begin{figure}
	\centering 
		\includegraphics [width = 8.5cm] {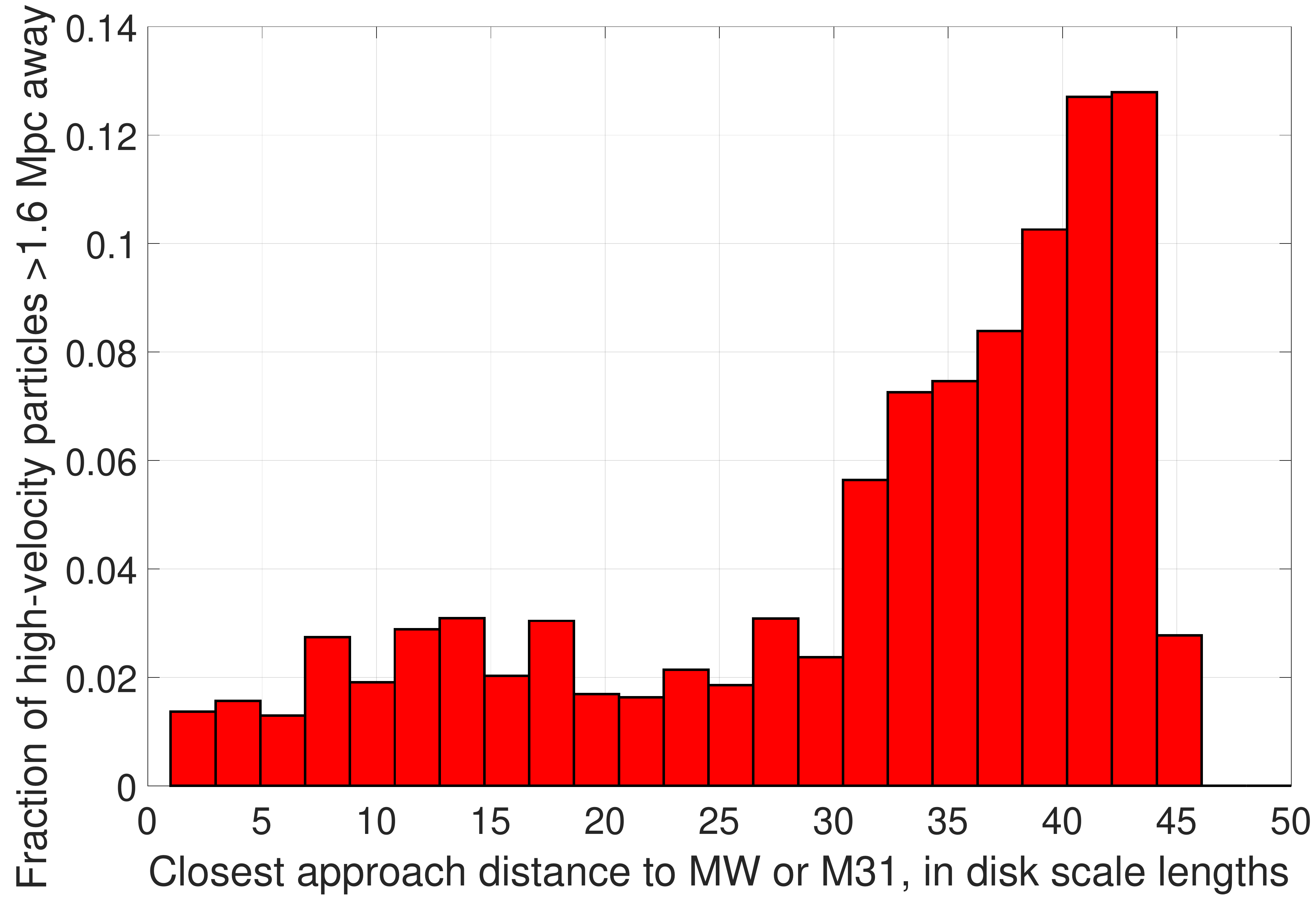}
		\caption{Histogram of the closest approach distances of high-velocity test particles (above dashed orange line in Figure \ref{Hubble_diagram_coloured_FL}) to the MW or M31 in units of their disk scale lengths. We show whichever of these quantities is smaller for any given test particle. It is evident that very few of these particles approached the MW or M31 so closely for the details of their mass distribution to become important. Each particle has been weighted according to the volume it represents in our 3D grid of initial conditions. This corresponds to the mass it represents if we assume that the LG initially had a uniform density except in appropriately sized spherical `holes' around the MW and M31 (Equation \ref{r_exc}).}
	\label{Closest_approach_distance_histogram}
\end{figure}

In Figure \ref{Hubble_diagram_coloured_FL}, we show the distances and radial velocities of test particles with respect to the LG barycentre, colour-coding them according to the orientation of their orbital plane. We quantify this based on the specific angular momentum $\bm{h}$, whose direction can readily be compared with the MW-M31 orbital pole $\widehat{\bm{h}}_{_{MW-M31}}$.
\begin{eqnarray}
	\bm{h} &\equiv & \bm{r} \times \dot{\bm{r}}	\\
	\cos \psi &\equiv & \widehat{\bm{h}} \cdot \widehat{\bm{h}}_{_{MW-M31}}
	\label{cos_psi_definition}
\end{eqnarray}

The initial positions of the particles span a grid in spherical polar co-ordinates. We do not show results for particles that pass within 15.4 kpc of the MW or 21.5 kpc of M31. In the real LG, such particles would likely have merged with the nearby galaxy.

For this work, the important feature is the upper branch of the Hubble diagram. Its upward slope arises because these particles must have passed close to the spacetime location of the MW-M31 encounter and gained a substantial amount of kinetic energy in what was essentially a 3-body interaction. Thus, for such particles to be further away from the LG now, they must have a larger outwards velocity. This will depend somewhat on when each particle approached whichever of the MW or M31 most strongly scattered that particle $-$ and thus the speed of the perturber at that time.

However, the precise encounter distance $b$ should not affect our results much because of the $r^{-1}$ gravity law (Equation \ref{Deep_MOND_limit}). Roughly speaking, doubling $b$ halves the strength of gravity but doubles the duration of the encounter, leaving the total impulse unchanged. This is probably why we found no clear correlation between how far particles were flung from the LG and how closely they approached the MW/M31. Thus, our results should not depend much on the minimum allowed encounter distance in our simulation or on the fact that the MW and M31 have finite extents and are not point masses. The important effect here is the time dependence of the gravitational field at the positions of the test particles which get flung out at high speed. Most of these never come that close to the MW or M31 because the commoner more distant encounters are just as effective at scattering test particles (Figure \ref{Closest_approach_distance_histogram}).

\begin{figure}
	\centering 
		\includegraphics [width = 8.5cm] {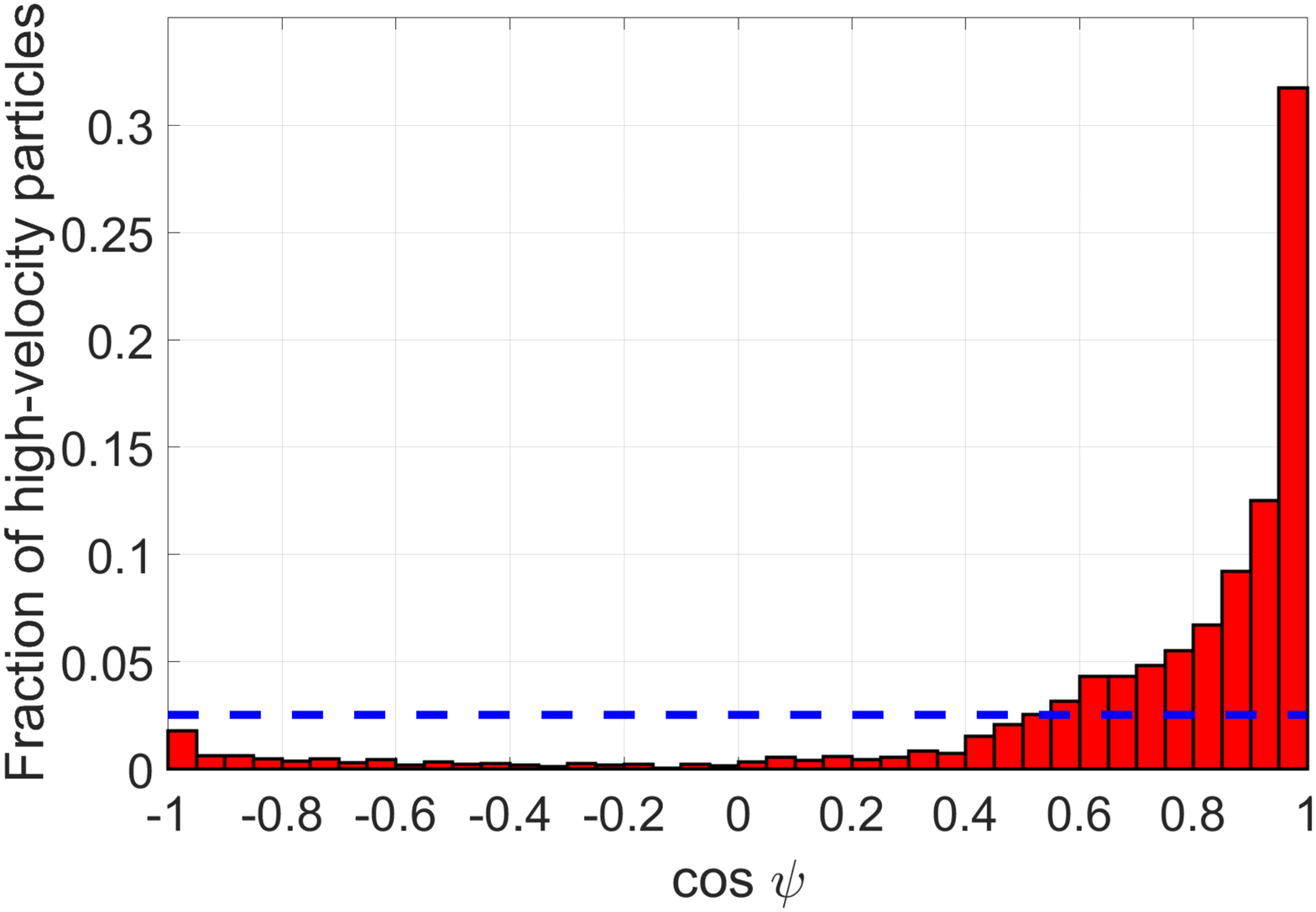}
		\caption{Histogram of $\cos \psi$ for all high-velocity test particles (above dashed orange line in Figure \ref{Hubble_diagram_coloured_FL}) beyond 1.6 Mpc from the LG barycentre in order to best correspond to actual HVGs (Figure \ref{Distance_GRV_correlation_3D_GA}). The dashed blue line indicates that 0.025 of the (weighted) particles should fall into each bin in $\cos \psi$ if their orbital poles were distributed isotropically.}
	\label{Cos_angle_histogram_1_6_Mpc}
\end{figure}

$\Lambda$CDM also allows slingshot encounters with the MW and M31, but their fairly slow motion means that they can only fling galaxies out to ${\ssim 1}$ Mpc from the LG, at which point the upper branch of the Hubble diagram simply stops \citep[][Figure 3]{Banik_Zhao_2016}. Even in more detailed cosmological simulations of $\Lambda$CDM that include encounters with satellites of MW and M31 analogues, dwarf galaxies do not get flung out beyond this distance \citep[][Figures 3 and 6]{Sales_2007}. For MOND, the corresponding limit is ${\ssim 2.5}$ Mpc due to the MW-M31 flyby, which therefore makes a dramatic difference to the Hubble diagram at distances of ${\ssim 1 - 2}$ Mpc (Figure \ref{Hubble_diagram_coloured_FL}). In this distance range, our simulation yields a bimodal distribution of radial velocities, with the HVGs corresponding to particles in the upper branch.

A pattern of this sort is apparent in the kinematics of the observed LG (Figure \ref{Distance_GRV_correlation_3D_GA}). Despite appearances, our simulation indicates that only ${\ssim 10\%}$ of the particles at these distances belong to the upper branch, roughly consistent with the frequency of HVGs in Section \ref{Sample_selection}. Other galaxies should lie below the Hubble flow due to the effect of gravity. In a MOND universe, they would probably lie closer to the Hubble flow than in our simulation due to the gravitational field of large scale structures on the LG. This external field effect weakens the gravity exerted by the MW and M31 at long range \citep[e.g.][]{Banik_Zhao_2015}. It is not caused by tides but arises because MOND gravity is non-linear in the matter distribution (Equation \ref{Simple_interpolating_function}).


\begin{figure}
	\centering 
		\includegraphics [width = 8.5cm] {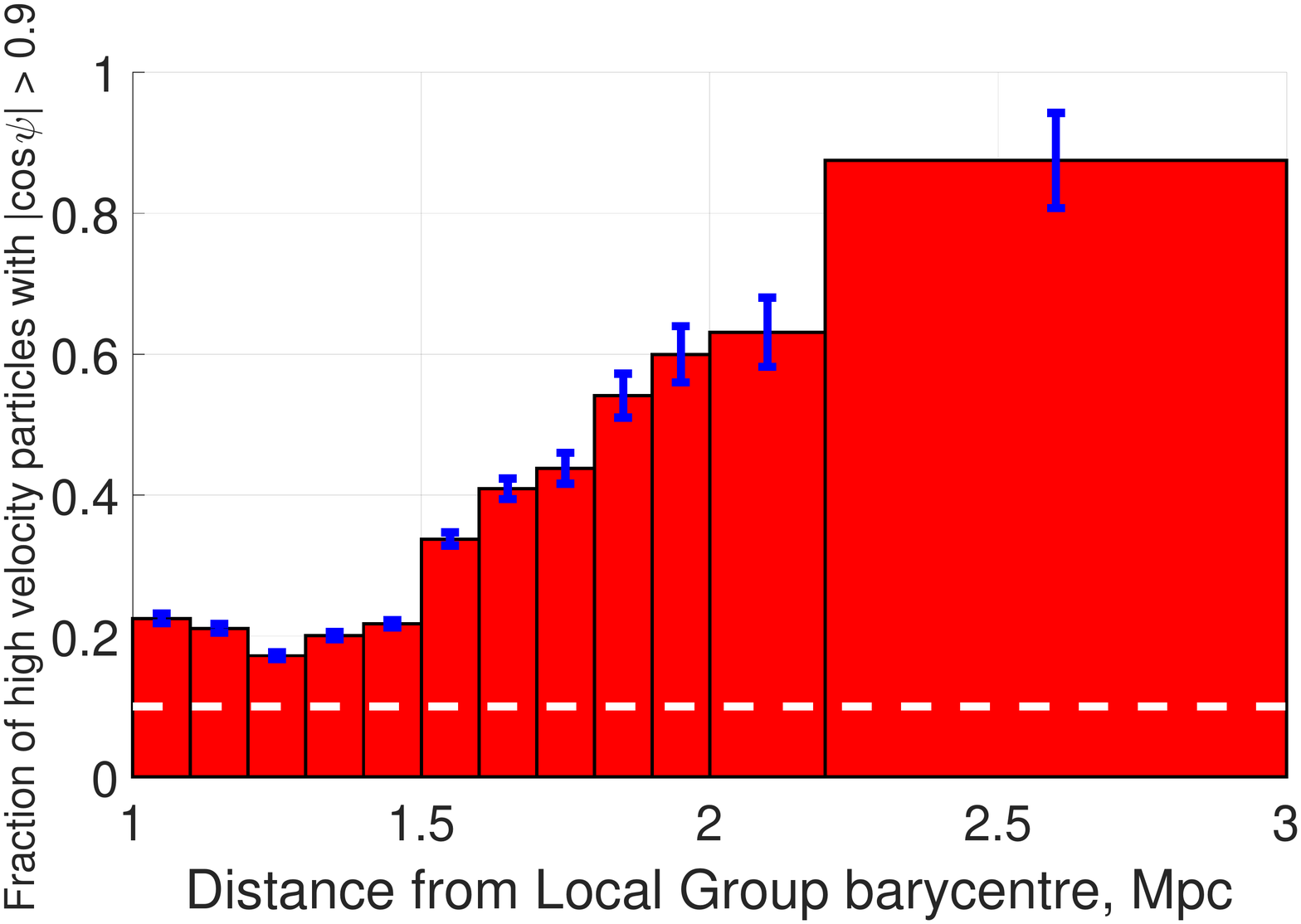}
		\caption{Fraction of high-velocity test particles (above dashed orange line in Figure \ref{Hubble_diagram_coloured_FL}) in each radial bin which have $\left| \cos \psi \right| > 0.9$. The dashed white line shows the expectation for an isotropic distribution (0.1). Notice how the high-velocity particles furthest from the LG now tend to have a more anisotropic distribution. Uncertainties are estimated using binomial statistics, though this is not totally accurate due to our statistical weighting scheme (see text). The result for the outermost radial bin is less reliable as we only have 24 particles in it, but other bins have at least 96 and should be quite reliable.}
	\label{Coplanar_fraction_results}
\end{figure}

In most parts of our simulated Hubble diagram, a moderate fraction of particles have orbits poorly aligned with the MW-M31 orbit (points coloured light blue, green or yellow in Figure \ref{Hubble_diagram_coloured_FL}). However, this is not true for the high-velocity branch at distances $\ga 1.5$ Mpc, which appears almost entirely dark red ($\cos \psi \approx 1$). This leads us to do a careful analysis of whether such particles really are distributed anisotropically.

Our initial grid of test particle positions is uniform in distance from the LG as well as in the spherical polar and azimuthal angles. Thus, each particle does not correspond to exactly the same volume/mass. We handle this by weighting each particle according to the volume it represents, which we find by integrating the usual spherical Jacobian factor over the range of initial co-ordinates covered by the particle\footnote{i.e. between the mid-points of the particle of interest and the particles at slightly smaller and larger polar angle etc.}.

We apply this weighting scheme to the high-velocity particles (above dashed orange line in Figure \ref{Hubble_diagram_coloured_FL}) to determine their correctly weighted distribution over $\cos \psi$ (Figure \ref{Cos_angle_histogram_1_6_Mpc}). If the particles have no preferred direction(s), then $\cos \psi$ should be distributed uniformly. Because we are investigating HVGs towards the edge of the LG, we also restrict to particles beyond 1.6 Mpc from its barycentre (as suggested by Figure \ref{Distance_GRV_correlation_3D_GA}).

A large proportion of high-velocity particles appear to have very high values of $\left| \cos \psi \right|$. To see how robust this is, we determine the (weighted) fraction of particles in different radial bins with $\left| \cos \psi \right| > 0.9$, our proxy for an orbital plane almost aligned with that of the MW and M31. We estimate uncertainties using binomial statistics, which is only approximately correct here due to our weighting procedure. Apart from the outermost radial bin, there should be enough simulated particles in each one to accurately estimate this fraction.

\begin{figure}
	\centering 
		\includegraphics [width = 8.5cm] {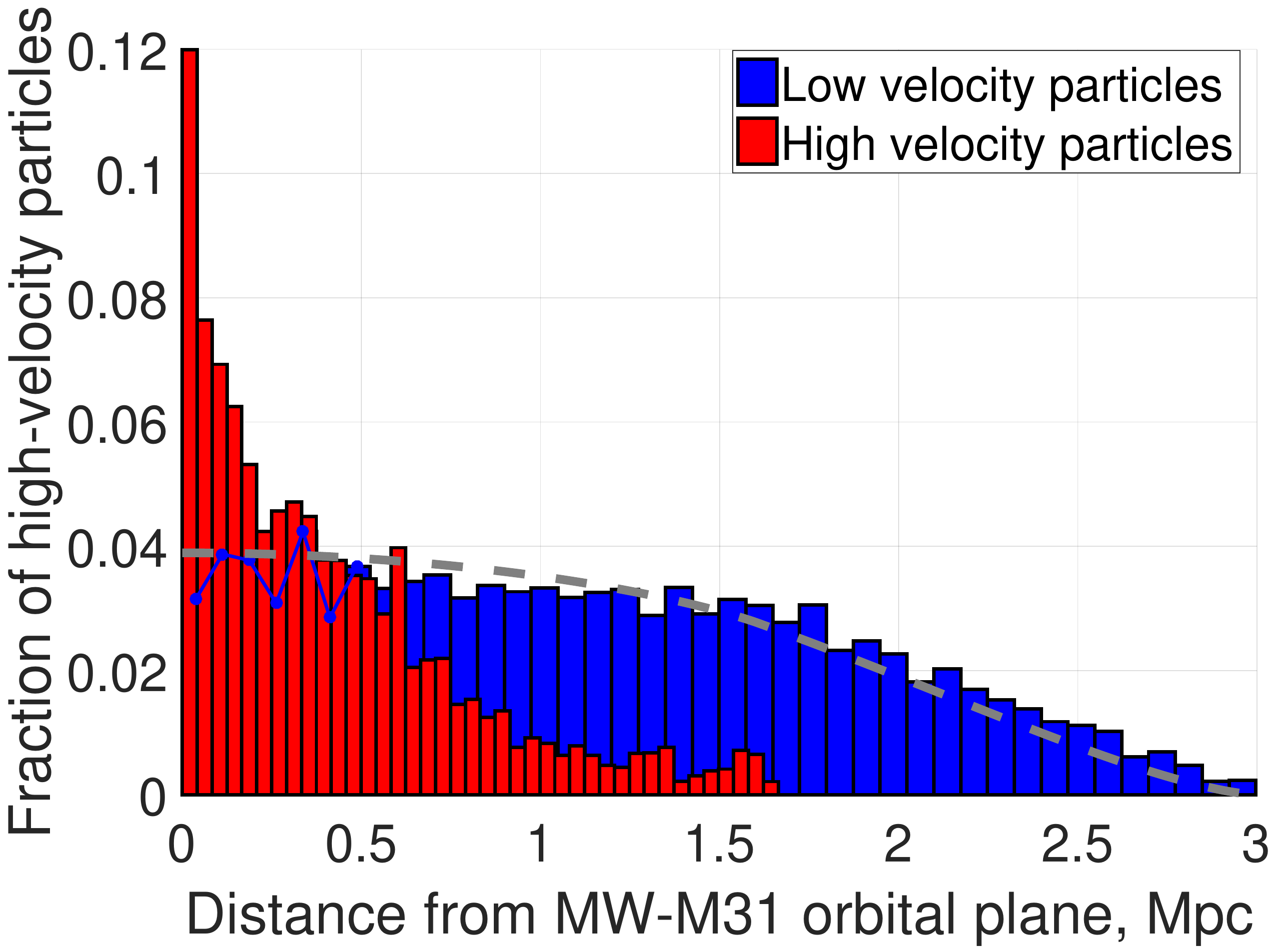}
		\caption{Histogram showing how far simulated particles are from the MW-M31 orbital plane. We only show particles currently at distances of ${1.6-3}$ Mpc from the LG and sort them according to whether they are in the high-velocity branch of the Hubble diagram (above dashed orange line in Figure \ref{Hubble_diagram_coloured_FL}). If they are, we show them as red. The remaining particles (shown in blue) are well described by an isotropic distribution (dashed grey line).}
	\label{z_distribution}
\end{figure}

Compared to an isotropic distribution, the fraction of nearly co-planar particles is very high (Figure \ref{Coplanar_fraction_results}). This demonstrates that dwarfs flung out furthest from the LG should be distributed very anisotropically in a MOND context. Moreover, the preferred plane should correspond to the MW-M31 orbital plane. Within this plane, the HVGs should mostly be co-rotating with respect to the MW-M31 orbit (Figure \ref{Cos_angle_histogram_1_6_Mpc}). However, it is not possible to test counter-rotation vs co-rotation at present due to a lack of accurate proper motions for the HVGs. This is why we focus on $\left| \cos \psi \right|$ rather than $\cos \psi$.

Gravitational slingshot interactions with the MW or M31 would be most efficient for particles flung out almost parallel to the motion of the perturber. Considering that the MW-M31 flyby occurred a fixed time in the past, these particles should currently be furthest away from the LG. Thus, it is not very surprising that the spatial distribution of such particles is highly flattened with respect to the MW-M31 orbital plane (Figure \ref{z_distribution}).

Although this scenario almost exclusively leads to HVGs co-rotating with respect to the MW-M31 orbit, that is not always the case. For a particle flung out on an almost radial orbit with respect to the LG, only a small torque is needed to reverse the direction of its angular momentum. This may explain why the high-velocity test particles with $\cos \psi \approx -1$ tend to have rather small angular momenta. \citet{Pawlowski_2011} suggested a similar mechanism to explain why some MW satellites like Sculptor are counter-rotating within the plane preferred by most remaining MW satellites \citep{Piatek_2006}.


\section{The Local Group in $\Lambda$CDM}

\subsection{Refining the $\Lambda$CDM model}
\label{Model_refinements}

To better identify which galaxies may have been flung out by a fast-moving MW/M31 in the way discussed in Section \ref{MOND_simulation}, we refine our previous $\Lambda$CDM dynamical model of the LG \citep{Banik_Zhao_2017}. In this work, we use an updated input catalogue. The main changes are a more accurate distance measurement to NGC 404 \citep{Dalcanton_2009} and to Leo P \citep{McQuinn_2015}. For NGC 4163, we use a less accurate distance of ${2.95 \pm 0.07}$ Mpc to bracket the range between the measurements of \citet{Dalcanton_2009} and \citet{Jacobs_2009}, both of which are based on data from the Hubble Space Telescope.

We also make improvements to the procedure used to find the best-fitting model parameters. The flatline level of the MW rotation curve is no longer assumed equal to its amplitude $v_{c, \odot}$ at the position of the Sun.\footnote{This is sometimes called the Local Standard of Rest (LSR).} We let the former vary with a prior of $205 \pm 10$ km/s \citep{McGaugh_2016_MW} while $v_{c, \odot}$ is fixed at 232.8 km/s \citep{McMillan_2017}. The time resolution is improved ${10 \times}$ so that 5000 steps are now used to cover the history of the Universe since redshift 9 (${a = 0.1}$), leading to a much better handling of close encounters.

The best-fitting solution is found by applying gradient descent to all model parameters, using a method similar to that described in Section \ref{Best_fitting_plane_finding}. To maximise the chance of matching observations, we run a grid search through the trajectories of all the dwarf galaxies, which are treated as test particles. Because the algorithm uses a least action method \citep{Shaya_2011}, trajectories are solved by relaxing an initial guess towards a solution that satisfies the equations of motion. The initial guess has the co-moving position varying linearly with $a$. The direction and magnitude of the present peculiar velocity of each dwarf are varied over a 3D grid of possibilities, giving the algorithm a much better chance of finding slingshot encounters that might otherwise get missed if the initial trajectory went nowhere near the spacetime location of the encounter. The issue of local but not global minima can always be solved with a grid search, which in this case is feasible for the dwarf galaxies because the trajectory of each one does not influence the gravitational field in the LG and thus the trajectory of anything else.\footnote{We did not do a full grid search as that would involve jointly varying trajectories of all 21 massive particles, impossible in a reasonable timeframe.}

As some improvements are indeed found in this way, we repeat the gradient descent stage and the grid search in an alternating manner until the algorithm converges in the sense that the grid search stops improving the agreement between model and observations. This process takes a few days and yields reliable trajectories for all simulated galaxies $-$ their present-day positions and velocities are almost perfectly recovered (maximum errors of 9 pc and 16 m/s, respectively) if we solve their trajectories forwards using a more traditional fourth-order Runge-Kutta method with $10\times$ finer resolution.

\begin{figure}
	\centering 
		\includegraphics [width = 8.5cm] {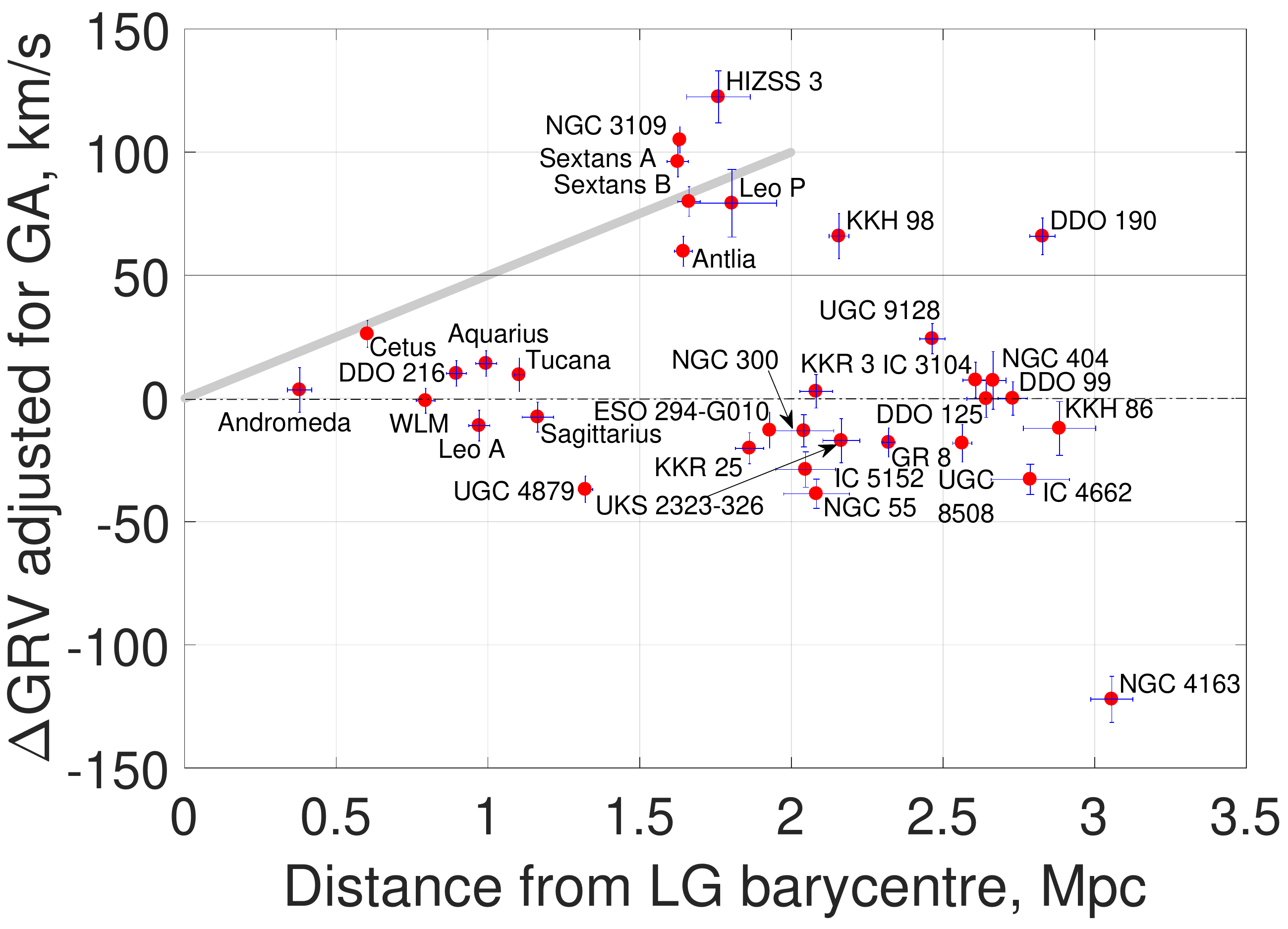}
		\caption{The deviation $\Delta GRV$ of each target galaxy from our best-fitting $\Lambda$CDM model is shown against its distance from the LG barycentre. If the model worked perfectly, then all galaxies would have ${\Delta GRV \equiv 0}$ (dot-dashed line) as model predictions are subtracted. Given likely model uncertainties of ${\ssim 25}$ km/s \citep{Aragon_Calvo_2011}, $\Lambda$CDM would thus find it difficult to explain galaxies with ${\Delta GRV > 50}$ km/s (above solid black gridline). In our MOND scenario of a past MW-M31 flyby, we expect the HVGs to broadly follow a trend of 50 km/s/Mpc (diagonal grey line) and to reach distances up to ${\ssim 2}$ Mpc (Figure \ref{Hubble_diagram_coloured_FL}).}
	\label{Distance_GRV_correlation_3D_GA}
\end{figure}

Using Equation 30 from \citet{Banik_Zhao_2017}, we adjust the predictions of this best-fitting model for the effect of tides raised on the LG by the Great Attractor. This only slightly affects our results, which are shown in Figure \ref{Distance_GRV_correlation_3D_GA}. The main difference from our previously published results \citep[][Figure 13]{Banik_Zhao_2017} is that Tucana is now consistent with $\Lambda$CDM expectations.

At distances ${\ga 1.5}$ Mpc, a bimodal distribution of $\Delta GRV$s is apparent, similar to that in our MOND simulation of the LG (Figure \ref{Hubble_diagram_coloured_FL}). Moreover, the galaxies in the lower branch predominantly have $\Delta GRV < 0$, perhaps a sign of the stronger gravity in MOND than in the $\Lambda$CDM model whose predictions have been subtracted.

\subsection{Selecting high-velocity galaxies}
\label{Sample_selection}

To find HVGs in the real LG, we compare the distances $d$ of our target galaxies from the LG barycentre with their $\Delta GRV \equiv GRV_{obs} - GRV_{model}$ relative to our best-fitting 3D dynamical model of the LG (Figure \ref{Distance_GRV_correlation_3D_GA}). We expect that these dwarf galaxies were flung out at high speed by the MW or M31, implying they passed close to the spacetime location of the MW-M31 flyby. Thus, such dwarfs should follow a $\Delta GRV \appropto d$ relation of the sort apparent in our MOND simulation of the LG (Figure \ref{Hubble_diagram_coloured_FL}). A relation like this is evident in Figure \ref{Distance_GRV_correlation_3D_GA}, where we have added a solid grey line at $u = 50$ km/s/Mpc to make it clearer. A radial velocity excess of this magnitude suggests that the MW-M31 flyby occurred $\sim \left(H_{_0} + u \right)^{-1} \approx 8$ Gyr ago. This is consistent with their expected orbital evolution in MOND (Figure \ref{MW_M31_trajectory_FL}).

A larger MW-M31 pericentre would not affect this conclusion much as a HVG would still require a similar velocity to reach its presently observed position from the spacetime location of the MW-M31 flyby, the timing of which is constrained observationally if we assume this event led to the formation of the MW thick disk \citep{Quillen_2001}. However, a weaker MW-M31 encounter would reduce the maximum distance at which we might expect to see a HVG.

As our $\Lambda$CDM-based model of the LG is not a perfect representation of a $\Lambda$CDM universe, we expect model uncertainties of $\ssim 25$ km/s based on how a LG analogue deviates from spherical symmetry in a detailed cosmological simulation of $\Lambda$CDM \citep{Aragon_Calvo_2011}.\footnote{This is discussed in more detail in Section 4 of \citet{Banik_Zhao_2016}.} Thus, we focus our attention on the galaxies with $\Delta GRV > 50$ km/s and following the $\Delta GRV \appropto d$ relation. This leads to the HVG sample in Table \ref{Planar_backsplash_galaxies}. 



The reasonably high $\Delta GRV$ of DDO 190 (${66 \pm 7}$ km/s) is still marginally compatible with our model if we assume a model uncertainty of $\ssim 25$ km/s. This is especially true when considering that the much larger distance of DDO 190 from the LG barycentre suggests that it should have a much higher $\Delta GRV$ if it really was flung out in the same way as e.g. NGC 3109. Thus, we do not consider DDO 190 as being a genuine HVG, even though its $\Delta GRV$ is slightly on the high side.

Although it would be quite normal to have one such instance amongst our 34 LG target galaxies of observations exceeding model predictions by ${2.6 \sigma}$ (probability $\approx 0.13$), a second such instance would be unexpected. Thus, it would be rare to observe $\Delta GRV$s as large as for DDO 190 and KKH 98 (${66 \pm 9}$ km/s) if we treat both as having normal kinematics in a $\Lambda$CDM context. Given that DDO 190 deviates very substantially from the $\Delta GRV \propto d$ relation typically followed by HVGs, this suggests that KKH 98 may be a HVG. Although it does not fit the $\Delta GRV \propto d$ relation perfectly, some scatter about this is expected because the LG is not spherically symmetric and is presently observed from an off-centre vantage point. Because most HVGs lie at rather similar angles to the MW-M31 line, we expect larger deviations from this relation for HVGs like KKH 98 which lie at a totally different angle (Figure \ref{Plane_position_Delta_GRV_GA}).

For the particular case of KKH 98, its position rather close to the MW-M31 line means that its $\Delta GRV$ would be more sensitive to where our model puts the centre of mass of the LG i.e. its preferred MW:M31 mass ratio. If the MW and M31 masses are not equal but only 0.3 of their total mass is in the MW (Section \ref{MOND_simulation}), then the LG barycentre would be shifted by ${\ssim 160}$ kpc towards M31 and thus by a similar amount towards KKH 98. This would put it closer to the LG barycentre. In the LG, the radial velocity rises with distance at a rate close to 100 km/s/Mpc due to the gravity of the MW and M31 \citep[][Figure 5]{Banik_Zhao_2017}. Thus, a galaxy 160 kpc closer to the LG barycentre should be receding away from it 16 km/s slower. Moreover, the MW would be moving 22 km/s faster towards M31 (and thus KKH 98) given the observed MW-M31 relative radial velocity \citep{M31_motion}. A more massive M31 would also be expected to reduce GRVs of objects in the general vicinity of KKH 98 compared to a situation where the MW and M31 have equal mass. Even without this dynamical effect, the kinematic effects alone would reduce the predicted GRV of KKH 98 by ${\ssim 40}$ km/s but would have a smaller effect on the other HVGs and DDO 190 as their sky positions are almost orthogonal to the MW-M31 line (Figure \ref{Plane_position_Delta_GRV_GA}). If this is correct, it explains why the $\Delta GRV$ of KKH 98 falls below the $\Delta GRV \propto d$ relation by about this much.

\begin{table}
 \begin{tabular}{ccc}
		\hline
		Galaxies included & Distance from MW-M31 & $\Delta GRV$, km/s \\
    in our plane fit & mid-point, Mpc & \\
		\hline
		Milky Way & $0.382 \pm 0.04$ & NA \\
		Andromeda & $0.382 \pm 0.04$ & $3.5 \pm 9.1$ \\ [5pt]
		Sextans A & $1.624 \pm 0.036$ & $96.1 \pm 6.3$ \\
		Sextans B & $1.661 \pm 0.037$ & $79.9 \pm 6.0$ \\
		NGC 3109 & $1.631 \pm 0.014$ & $105.0 \pm 5.3$ \\
		Antlia & $1.642 \pm 0.030$ & $59.7 \pm 6.1$ \\
		Leo P & $~1.80 \pm 0.15$ & $79 \pm 14$\\ [5pt]
		KKH 98 & $2.160 \pm 0.033$ & $65.5 \pm 9.1$\\
		\hline
 \end{tabular}
 \caption{Galaxies considered when finding the plane best fitting the high $\Delta GRV$ galaxies in our sample, which we select based on Figure \ref{Distance_GRV_correlation_3D_GA}.}
 \label{Planar_backsplash_galaxies}
\end{table}

Our scenario implies that any plane passing close to most of the HVGs should also pass close to the MW and M31. Thus, we apply our plane-fitting algorithm (Section \ref{Best_fitting_plane_finding}) to the galaxies listed in Table \ref{Planar_backsplash_galaxies}, always including the MW and M31 in our sample. We exclude HIZSS 3 despite the fact that it should be treated as a HVG because a much thinner plane is obtained without it (Figure \ref{Plane_offset_Delta_GRV_GA}). Naturally, it would be more common to find a sample of galaxies with an anisotropic distribution if one of them can be removed arbitrarily with the explicit intention of making the remaining ones have a more anisotropic distribution. We account for this using the method in Section \ref{Statistical_analysis_method}. There are also good observational reasons for excluding HIZSS 3 from our analysis, in which case this `look elsewhere' effect should not be considered (Section \ref{No_HIZSS}).

\section{Analysing The Local Group}
\label{Method}

\subsection{Finding the best-fitting plane}
\label{Best_fitting_plane_finding}

To quantify whether a set of galaxies is distributed anisotropically, we need to define a measure of anisotropy and determine how unusual its value is. The statistic we will use is $z_{_{rms}}$, the root mean square (rms) of the minimum distances between the galaxies we consider and the best-fitting plane through them (i.e. the one that minimises $z_{_{rms}}$). With respect to a plane having normal $\widehat{\bm{n}}$ and containing the vector $\bm{r}_{_0}$, the vertical dispersion is
\begin{eqnarray}
	\label{z_rms}
	{z_{_{rms}}}^2 ~&=&~ \frac{1}{N}\sum_{i = 1}^{N} \left[ \left( \bm{r}_{_i} - \bm{r}_{_0} \right) \cdot \widehat{\bm{n}} \right]^2 \\
	 ~&=&~ \widehat{\bm{n}} \cdot \left(\mathbf{I} \widehat{\bm{n}} \right) ~~\text{ where} \\
	\mathbf{I}_{jk} ~&\equiv &~ \frac{1}{N}\sum_{i = 1}^{N} \left( \bm{r}_{_i} - \bm{r}_{_0} \right)_j \left( \bm{r}_{_i} - \bm{r}_{_0} \right)_k
\end{eqnarray}

The galaxies are at heliocentric positions $\bm{r}_{_i}$. The minimum of $z_{_{rms}}$ is attained when $\bm{r}_{_0} = \frac{1}{N} \sum_{i = 1}^{N} \bm{r}_{_i}$, corresponding to the geometric centre of the $N$ galaxies to which we are trying to fit a plane. We find the best-fitting orientation $\widehat{\bm{n}}$ using a gradient descent method \citep[e.g.][]{Fletcher_1963}. The gradient of ${z_{_{rms}}}^2$ with respect to $\widehat{\bm{n}}$ is $\frac{2}{N}\left( \mathbf{I} \widehat{\bm{n}} \right)$ less the component of this parallel to $\widehat{\bm{n}}$. At the minimum of $z_{_{rms}}$, its gradient vanishes, implying that $\widehat{\bm{n}}$ is an eigenvector of the inertia tensor $\mathbf{I}$ corresponding to its minimum eigenvalue. This provides a non-iterative way of minimising $z_{_{rms}}$, taking advantage of the characteristic polynomial of $\mathbf{I}$ being a cubic whose roots can be found analytically. However, we find that this approach is slower than gradient descent, a much more general method which we also used in Section \ref{Model_refinements}.

We minimise issues of local minima by starting the gradient descent based on whichever $\widehat{\bm{n}}$ yields the smallest $z_{_{rms}}$ in a low resolution grid of possible directions for $\widehat{\bm{n}}$. Once the angular step size is below ${0.006^\circ}$, we stop doing further iterations. As well as ensuring that our algorithm always converges in this sense, we also verify it using mock data designed to lie close to a plane with known $\widehat{\bm{n}}$ and $z_{_{rms}}$. We are always able to accurately recover their input values.

\subsection{Statistical analysis}
\label{Statistical_analysis_method}

\begin{table}
 \begin{tabular}{llll}
\hline
  Quantity & Full sample & Without Antlia\\ 
  \hline
  Galaxies in plane & 8 & 7 \\ [5pt]
  Normal to plane of & \multirow{2}{*}{$\begin{bmatrix} 204.4^\circ \\ -30.1^\circ \end{bmatrix}$} & \multirow{2}{*}{$\begin{bmatrix} 206.6^\circ \\ -31.8^\circ \end{bmatrix}$} \\
  high $\Delta GRV$ galaxies  &  \\ [5pt]
  rms plane width, kpc & 101.1 & 101.9 \\
  Aspect ratio (Eq. \ref{Aspect_ratio}) & 0.0763 & 0.0750 \\ [5pt]
  MW offset from plane & 224.7 & 195.4 \\
  M31 offset from plane & -0.6 & -12.8 \\
	Angle of MW-M31  & \multirow{2}{*}{$16.2^\circ$} & \multirow{2}{*}{$14.9^\circ$} \\
	line with plane & & \\
  \hline
 \end{tabular} 
 \caption{Information about the plane best fitting the galaxies listed in Table \ref{Planar_backsplash_galaxies}, with distances in kpc and plane normal direction in Galactic co-ordinates (latitude last). The last column shows how our results change if Antlia is removed from our sample as it could be a satellite of NGC 3109 \citep{Van_den_Bergh_1999}. The effect on our statistical analysis is described in Section \ref{Excluding_Antlia}.}
 \label{Plane_parameters} 
\end{table}

\begin{figure}
	\centering 
		\includegraphics [width = 8.5cm] {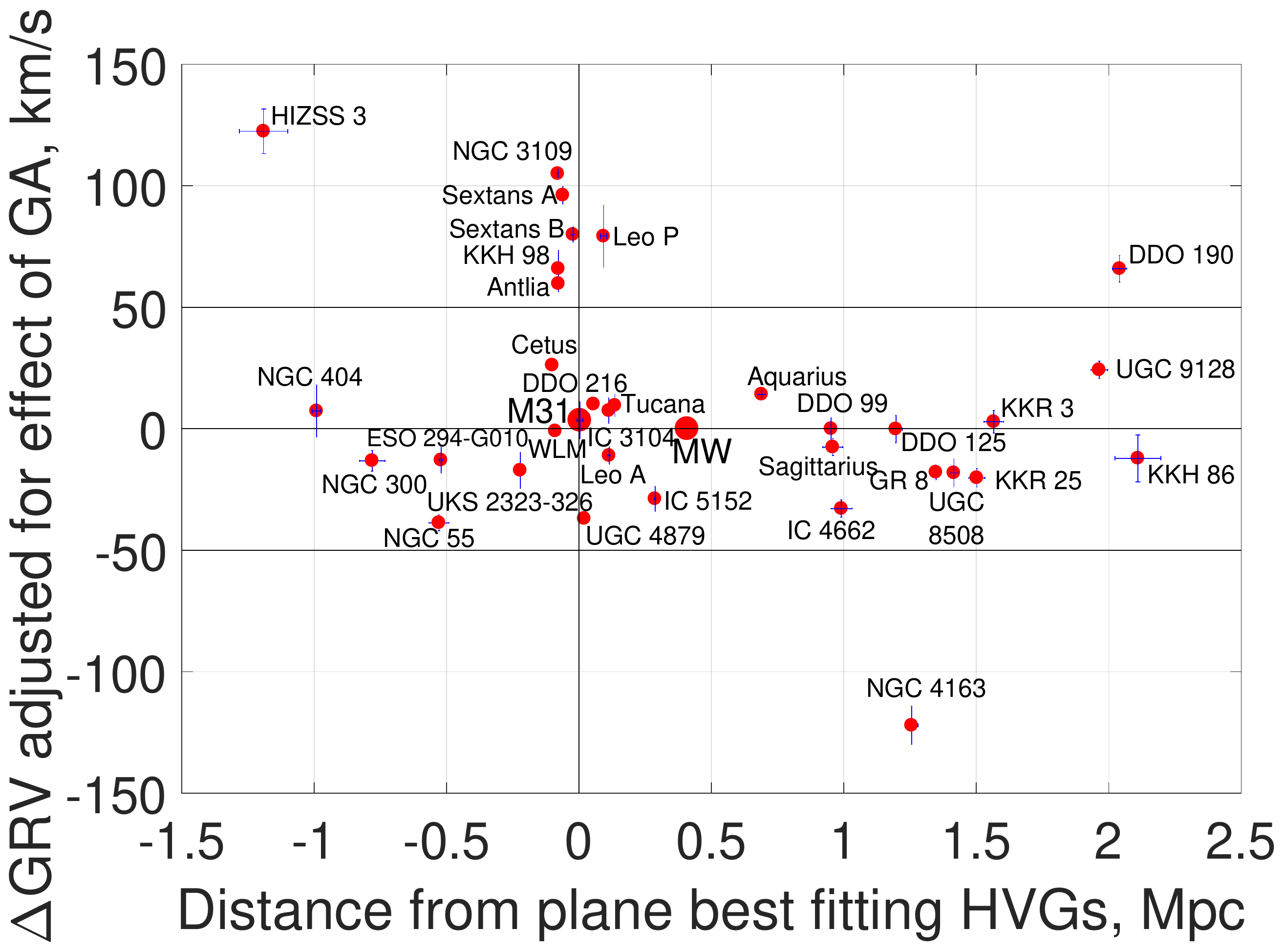}
	\caption{$\Delta GRV$s of target galaxies are shown against their offsets from the best-fitting plane through the ones with the largest $\Delta GRV$s except HIZSS 3 (parameters of this plane given in central column of Table \ref{Plane_parameters}). By definition, $\Lambda$CDM predicts $\Delta GRV = 0$ (central horizontal line) with an uncertainty of $\ssim 25$ km/s \citep{Aragon_Calvo_2011}. Thus, the model can't easily explain galaxies with ${\Delta GRV > 50}$ km/s (above upper horizontal line). Most of these galaxies lie very close to a plane (near vertical gridline), unlike the rest of our sample. For the MW, the concept of a $\Delta$GRV is meaningless, so we show this as 0.}
	\label{Plane_offset_Delta_GRV_GA}
\end{figure}

The MW, M31 and all but one of the HVGs lie close to a plane (Figure \ref{Plane_offset_Delta_GRV_GA}). We need to reflect this when determining the likelihood of $z_{_{rms}}$ being as low as the observed 101 kpc. To see if this is consistent with isotropy, we conduct a series of Monte Carlo (MC) trials in which we randomise the directions to these galaxies and recompute $z_{_{rms}}$. Thus, the probability distribution of the Galactic longitude $l$ is uniform while that of the Galactic latitude $b$ is
\begin{eqnarray}
	P(b)~db ~=~ \frac{1}{2}\cos b~db
\end{eqnarray}

To mimic uncertainties in measured distances to LG galaxies, we randomly vary their heliocentric distances using Gaussian distributions of the corresponding widths. In the very rare cases where this yields a negative distance, we set this to 0.

To account for HIZSS 3 being excluded from our plane fit despite its high $\Delta GRV$, we use the procedure described in Section \ref{Best_fitting_plane_finding} to find the best-fitting plane through every combination of all HVGs but one as well as the MW and M31.\footnote{It would not make sense for the major LG galaxies to lie far from this plane as it should be their mutual orbital plane.} The combination yielding the lowest $z_{_{rms}}$ is considered the analogue of the observed HVG system less HIZSS 3 for that particular randomly generated mock catalogue. In Section \ref{Using_aspect_ratio}, we perform calculations where we select the combination yielding the lowest aspect ratio $A$ rather than $z_{_{rms}}$.
\begin{eqnarray}
	\label{Aspect_ratio}
	A ~&\equiv&~ \frac{z_{_{rms}}}{\sqrt{{{r_{_{rms}}}^2 - {z_{_{rms}}}^2}}} ~~~\text{  where} \\
	{r_{_{rms}}}^2 ~&\equiv&~ \frac{1}{N} \sum_{i = 1}^{N} \left| \bm{r}_{_i} - \bm{r}_{_0} \right|^2 ~=~ Trace\left( \mathbf{I} \right)
\end{eqnarray}

$r_{_{rms}}$ is the rms distance of the galaxies from their geometric centre $\bm{r}_{_0}$. To get the rms extent of the system after projection into the best-fitting plane, we need to subtract $z_{_{rms}}$ in quadrature. Dividing $z_{_{rms}}$ by the result then gives a measure of the typical `vertical' extent of galaxies out of this plane relative to their `horizontal' extent within it. We would obtain identical probabilities for the observed situation if we defined $A$ as $\frac{z_{_{rms}}}{r_{_{rms}}}$ instead, as long as it is defined in the same way for the actual HVGs and the mock sample in each MC trial (Section \ref{Criteria_definitions}). This is because $A$ is a monotonic function of $\frac{z_{_{rms}}}{r_{_{rms}}}$ with either definition.

The major LG galaxies along with the HVGs except HIZSS 3 (full list in Table \ref{Planar_backsplash_galaxies}) define a rather thin plane whose parameters are given in Table \ref{Plane_parameters}. This allows us to compare the $\Delta GRV$ of each galaxy\footnote{adjusted for the Great Attractor using Equation 30 of \citet{Banik_Zhao_2017}} with its minimum distance from this plane. The galaxies in our full sample have a wide range of positions relative to it, with a similar number on either side (Figure \ref{Plane_offset_Delta_GRV_GA}). However, the HVGs tend to lie very close to it. The only exception is HIZSS 3, justifying our decision not to consider it when defining the HVG plane. In any case, the observations for HIZSS 3 are rather insecure (Section \ref{No_HIZSS}).

In Table \ref{Criteria_definitions}, we give the criteria which we use to determine whether the distribution of HVGs in a MC trial is analogous to their observed distribution. We choose these criteria so that they should be satisfied if the LG behaves similarly to our MOND simulation of it (Section \ref{MOND_simulation}). Depending on which measure of anisotropy is used, we consider one of the first two criteria alongside both of the others.

In Section \ref{Orbital_plane_finding}, we used a toy model to find the MW-M31 orbital pole $\widehat{\bm h}$ leading to a past close encounter between these galaxies in the most favourable orientation for the formation of their satellite planes. One might expect the HVGs to define this orbital plane. Thus, it is interesting that there is only a ${\ssim 19^\circ}$ angle between the plane normal defined by the HVGs and our expected direction for $\widehat{\bm h}$.

Another requirement of our model is that the HVG plane should intersect the MW-M31 barycentre, which we take to be $0.3$ of the way from M31 towards the MW for reasons discussed at the start of Section \ref{MOND_governing_equations}.\footnote{Even in $\Lambda$CDM, it is likely that M31 has a higher mass than the MW \citep{Jorge_2014, Banik_Zhao_2016}.} This puts the MW-M31 barycentre 67 kpc from the best-fitting plane, a rather small offset from a plane with a radial extent of ${\ssim 1.3}$ Mpc.

\begin{table}
 \begin{tabular}{ll}
		\hline
		Criterion & Meaning\\
		\hline
		Plane & There must be a plane of HVGs with rms\\
		thickness & thickness (Equation \ref{z_rms}) below that observed\\ [5pt]
		Aspect & There must be a plane of HVGs with aspect \\
		ratio & ratio (Equation \ref{Aspect_ratio}) below that observed\\ [5pt]
		Barycentre & Barycentre of MW and M31 (assuming\\
		offset &30\% of total mass in MW) closer to\\
		& plane than observed situation\\ [5pt]
		Direction & Normal to HVG plane closer to expected\\
		& direction (Section \ref{Orbital_plane_finding}) than observed\\
		\hline
 \end{tabular} 
 \caption{Criteria used to judge whether a randomly generated population of galaxies is analogous to the observed population of HVGs. Only one of the first two criteria is used at a time. Note that the criteria are not all independent. For example, as the MW and M31 positions are fixed and our plane fitting procedure always considers them, the thinnest planes are likely to be obtained when these galaxies are close to the plane best fitting the HVGs. This makes it more likely that the `barycentre offset' criterion is satisfied (see corners of Table \ref{Probability_table_criteria_combinations}).}
 \label{Criteria_definitions}
\end{table}

\section{Results}
\label{Results}

Applying the criteria defined in Table \ref{Criteria_definitions} to 20 million MC trials based on our nominal sample of HVGs (Table \ref{Planar_backsplash_galaxies}), we obtain the results shown in Table \ref{Probability_table_criteria_combinations}. The uncertainties are found by repeating the MC trial using 4 different seeds for the random number generator, with each seed used for ${5 \times 10^6}$ trials.\footnote{We use the rng(`shuffle') command in \textsc{matlab}$^\text{\textregistered}$ to initialise its random number generator based on the date and time. We verify that different runs give slightly different results, which is inevitable for runs started at least a few seconds apart.} The variance between the results gives an indication of the uncertainty in our final result, which is a simple mean. We also estimate the error using binomial statistics. Our final error estimate is based on whichever method suggests a higher uncertainty. Usually, this is the method involving comparing results obtained using different random number seeds. In all cases, we are able to constrain the proportion of MC trials matching our criteria to within a few percent of its most likely value.

\begin{table}
 \begin{tabular}{cccc}
\hline
 & Thickness & Direction & Barycentre \\
 & & & offset \\ \hline
Thickness & $4.6 \pm 0.3$ &  & \\
Direction & $2.3 \pm 0.1$ & $417.4 \pm 0.5$ & \\
Barycentre offset & $2.4 \pm 0.2$ & $81.1 \pm 1.3$ & $181.8 \pm 2.5$ \\
\hline
 \end{tabular} 
 \caption{Monte Carlo trial-based probabilities in parts per thousand (\perthousand) of the HVGs (all galaxies in Table \ref{Planar_backsplash_galaxies}) matching various combinations of the criteria defined in Table \ref{Criteria_definitions}. These criteria are used to determine if a mock system of HVGs is analogous to the observed system, using the method outlined in Section \ref{Statistical_analysis_method}. When the same criterion appears in both the row and column headings, the result is the probability of matching that criterion alone, regardless of the others. The probability of all three criteria being met simultaneously is $1.48 \pm 0.10\perthousand$, which corresponds to the first row of Table \ref{Probability_table}.}
 \label{Probability_table_criteria_combinations}
\end{table}

The direction criterion proved the least problematic due to the rather wide range of allowed orientations of the plane best fitting the mock galaxies. This criterion was met ${\ssim 417\perthousand}$ of the time.

The plane of HVGs is offset from the MW-M31 barycentre by 67 kpc, which is rather small considering the extent of the HVG plane ($\ssim 1$ Mpc). Thus, the `barycentre offset' criterion in Table \ref{Criteria_definitions} is only met around ${182\perthousand}$ of the time.

\begin{table}
 \begin{tabular}{lll}
	\hline
  Investigation & Sample & Probability (\perthousand)\\ 
   &  &   \\ 
  \hline
  Nominal (physical thickness) & All & $1.48 \pm 0.10$ \\
  $\widehat{\bm h}$ rotated $5^\circ$ south ($\theta = 75^\circ$) & All & $1.51 \pm 0.10$ \\
  Distances fixed & All & $1.45 \pm 0.01$ \\
  Nominal & \st{HIZSS 3} & $0.41 \pm 0.02$ \\
  Nominal & \st{Antlia} & $5.17 \pm 0.36$ \\
	Aspect ratio & All & $1.62 \pm 0.01$ \\
  Aspect ratio & \st{Antlia} & $5.35 \pm 0.02$ \\
  \hline
 \end{tabular} 
 \caption{How our results depend on various model assumptions. The final column shows the probability of a MC trial satisfying the criteria given in Table \ref{Criteria_definitions} based on randomising the directions towards the HVGs in Table \ref{Planar_backsplash_galaxies}. Galaxies whose names have been crossed out are excluded from our sample in that particular investigation, with the nominal sample corresponding to the `full sample' column of Table \ref{Plane_parameters}. Unless stated otherwise, we use the `plane thickness' criterion and a MW-M31 orbital plane corresponding to $\theta = 70^\circ$ in Figure \ref{Flyby_geometry_grid}, with 0.3 of the total MW and M31 mass assumed to be in the MW (Section \ref{MOND_governing_equations}.}
 \label{Probability_table}
\end{table}

By far the most important criterion is the requirement that all but one mock HVG define a plane with rms thickness smaller than observed. This criterion is met only ${5.2 \pm 0.2\perthousand}$ of the time. Consequently, it is very unlikely (probability ${1.48 \pm 0.10\perthousand}$) that all three criteria are satisfied simultaneously.

To check how various assumptions affect our results, we conduct several versions of our statistical analysis. We describe these variations next and summarise our findings in Table \ref{Probability_table}.


\subsection{Altering the expected MW-M31 orbital plane}

In Section \ref{Orbital_plane_finding}, we used a toy model for the origin of the MW and M31 satellite planes to estimate $\widehat{\bm h}$, the direction of the MW-M31 orbital angular momentum. We parametrised this using the angle $\theta$, which we take to be $70^\circ$. The actual value may well be different, making it important to know if this affects our conclusions. To this end, we repeat the analysis shown in Table \ref{Probability_table_criteria_combinations} with $\theta = 75^\circ$. A slightly different expected $\widehat{\bm h}$ alters the range of `allowed' orientations for the best-fitting plane. However, the results hardly change (Table \ref{Probability_table}).

This is because a fairly wide range of plane normal directions are allowed such that this is not the main difficulty with matching the observed situation (Table \ref{Probability_table_criteria_combinations}). Instead, the difficulty lies in obtaining a plane as thin as the observed 101 kpc despite the HVGs having much larger distances (Table \ref{Planar_backsplash_galaxies}). Thus, in this contribution, we use $\theta = 70^\circ$ as we consider this to be more realistic. Although this does not much affect our results, a lower value for $\theta$ leads to a greater torque on the M31 disk, which is otherwise quite small. This is evident from our toy model preferring a solution close to the gap in Figure \ref{Flyby_geometry_grid} arising from the MW-M31 line at their closest approach lying almost within the M31 disk plane (Equation \ref{h_limits}). Thus, we think it will be easier for a MOND simulation to reproduce the observed properties of the MW and M31 disks and satellite planes using a slightly smaller value of $\theta$ or equivalently with $\widehat{\bm h}$ pointing slightly closer to the Galactic Equator.


\subsection{Using fixed distances}

Our analysis randomises both the directions towards the HVGs as well as their distances. The latter are drawn from a Gaussian distribution corresponding to the relevant measurement and its uncertainty, with any negative mock distances converted to 0. In the same way, we also vary the heliocentric distances to M31 \citep[${783 \pm 25}$ kpc,][]{McConnachie_2012} and to the centre of our own Galaxy \citep[${8.20 \pm 0.09}$ kpc,][]{McMillan_2017}. To see how this affects our results, we repeat our analysis using fixed distances to all galaxies in our sample. This has only a minor impact on our results, with the probability of matching all criteria falling from ${1.48 \pm 0.10 \perthousand}$ to ${1.45 \pm 0.01 \perthousand}$. This is consistent with no change, suggesting that distances to LG galaxies are known accurately enough for our analysis.

\subsection{Excluding low Galactic latitudes}
\label{No_HIZSS}

\begin{figure}
	\centering 
		\includegraphics [width = 8.5cm] {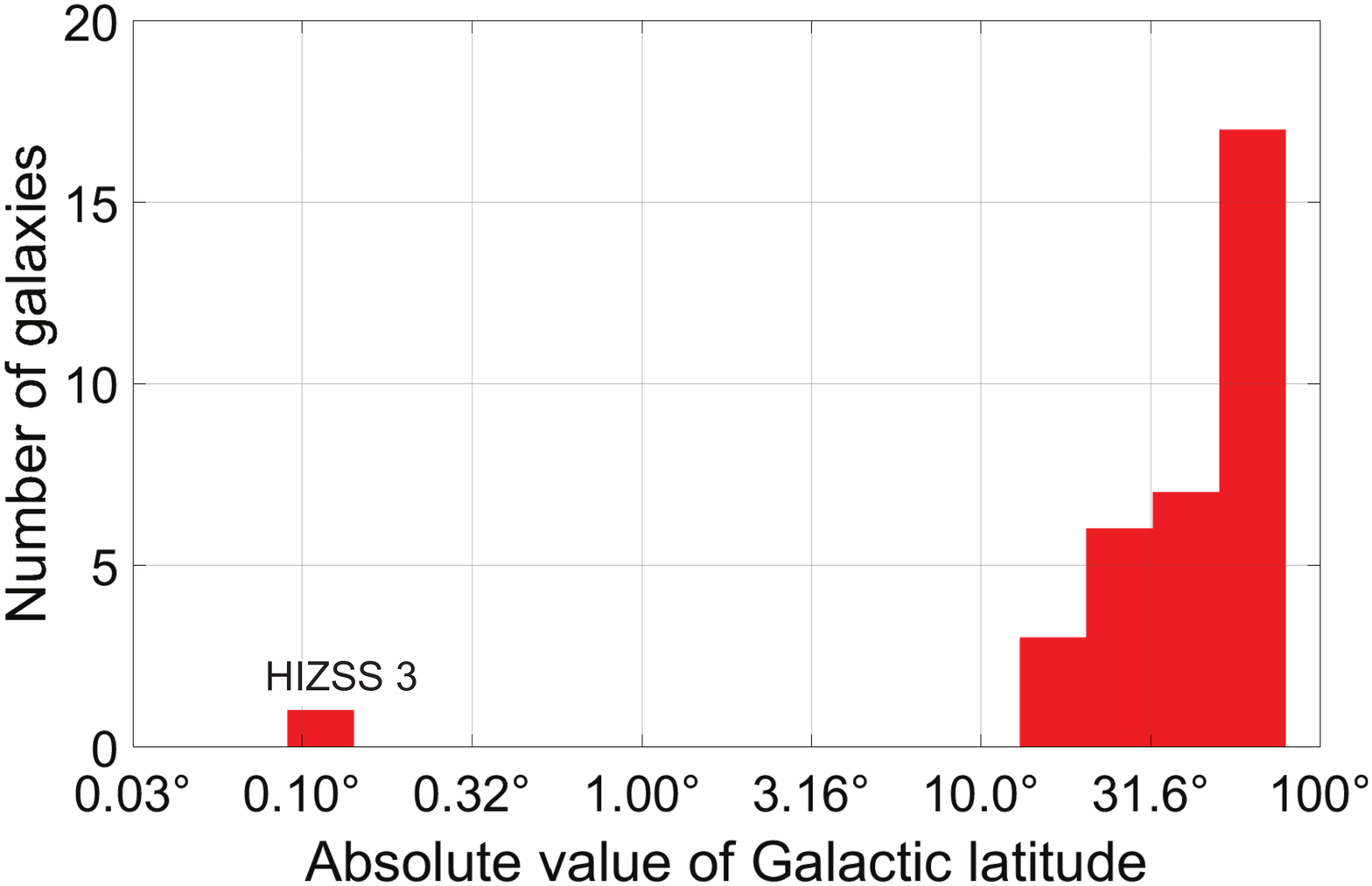}
	\caption{Histogram showing the Galactic latitudes of all galaxies in our sample (not just HVGs). Notice the very small value for HIZSS 3 \citep[${0.09^\circ}$,][]{Massey_2003}.}
	\label{Histogram_latitudes}
\end{figure}

We do not expect that \emph{all} HVGs necessarily lie close to a plane (Figure \ref{z_distribution}). Thus, it is not too surprising that one of them (HIZSS 3) lies far outside the plane defined by the others (Figure \ref{Plane_offset_Delta_GRV_GA}). However, it is also possible that there are observational issues for HIZSS 3 due to its extremely low Galactic latitude \citep[$b = {0.09^\circ}$,][]{Massey_2003}. It is readily apparent that no other galaxy in our entire sample (not just the HVGs) have sky positions closer to the Galactic disk (Figure \ref{Histogram_latitudes}).

The low Galactic latitude of HIZSS 3 may reduce the accuracy of its distance and/or radial velocity measurement. In particular, its tip of the red giant branch-based distance \citep{Silva_2005} seems rather insecure as this feature on its colour-magnitude diagram (CMD) is not very well defined (see their Figure 6). It is based on only a small number of stars in images suffering from a rather high contamination fraction by foreground MW stars due to the low $\left| b \right|$.

The authors also mention possible complications in the dust correction due to only reliably knowing this for the central part of HIZSS 3 but needing to use imaging over a much larger area of it in order to get enough stars in its CMD. In particular, the distance estimate would be higher if its central regions were particularly dusty, whereas a uniform dust correction was assumed. A more distant galaxy would have a lower $\Delta GRV$. This could be clarified with an updated distance to HIZSS 3, but unfortunately we only have one measurement that is now more than a decade old.

As well as contamination by foreground stars, Galactic dust is a major issue at very low Galactic latitudes. Due to the large amount of dust along such lines of sight, it becomes crucial to correct for this accurately. The foreground extinction measure $E \left( B - V \right)$ was estimated at 1.32 by \citet{Silva_2005}. However, more recent work suggests an extinction of only 0.88 \citep{Schlafly_2011}, with an uncertainty close to 0.1 in both cases. Reducing the extinction towards an astrophysical object increases its apparent luminosity. For consistency with observations, it must be further away than previously thought, in this case by ${\ssim 0.4}$ Mpc. This would increase the predicted GRV of HIZSS 3 by a substantial amount. A look at Figure 5 of \citet{Banik_Zhao_2017} suggests that the increase would be ${\ssim 50-100}$ km/s, perhaps resolving the discrepancy between the observed GRV of HIZSS 3 and that expected in our model. If this is correct, then HIZSS 3 is not a HVG.

\citet{Silva_2005} often referred to a basic (1D) dynamical model of the LG combined with the redshift of HIZSS 3 to try and justify their distance estimate. However, the law of gravity governing the LG and its past dynamical history are presently unclear, in particular whether there was a past close MW-M31 encounter \citep{Zhao_2013}. This makes it important to obtain redshift-independent distances. If this is not possible, then we note that HIZSS 3 is likely not a HVG as only a minority of galaxies are (e.g. if we use a threshold of $\Delta GRV = 50$ km/s in Figure \ref{Distance_GRV_correlation_3D_GA}). This is possible if it is further away than we assumed, as seems likely.

Apart from HIZSS 3, the galaxy in our sample with the lowest $\left| b \right|$ is the Sagittarius dwarf irregular galaxy \citep[$b = -16.3^\circ$,][]{Longmore_1978}. This suggests that observations at lower $\left| b \right|$ are difficult. Thus, we repeat our analysis with the directions towards the HVGs randomised but restricted such that $\left| b \right| \geq 15^\circ$. We consider this a plausible range of `unobservable' Galactic latitudes as a limit of $19.47^\circ$ was used by \citet{Cautun_2015} when dealing with MW satellites, though \citet{Lopez_Corredoira_2016} considered this a bit too high.

The major consequence of such a restriction is that HIZSS 3 must be removed from our sample. We think this is reasonable given the above reasons for why its distance measurement seems rather unreliable. However, no other galaxy is removed from our sample (Figure \ref{Histogram_latitudes}). This leads to \emph{all} HVGs lying very close to a plane, greatly reducing the probability of obtaining a situation matching the observed one (probability decreases from ${1.48 \pm 0.10\perthousand}$ to ${0.41 \pm 0.02\perthousand}$). This is mainly because the chance of getting a plane as thin as observed drops from ${4.6 \pm 0.3 \perthousand}$ to ${0.85 \pm 0.03 \perthousand}$.

\subsection{Excluding Antlia}
\label{Excluding_Antlia}

We have treated all HVGs as independent LG dwarf galaxies. In particular, we assumed that NGC 3109 and Antlia are unrelated objects. However, they may be gravitationally bound \citep{Van_den_Bergh_1999}. There are indications that they have recently interacted, based on observations of both NGC 3109 \citep{Barnes_2001} and Antlia \citep{Penny_2012}. This is more likely if Antlia is a satellite of NGC 3109. The ${41 \pm 1}$ km/s difference in their radial velocities \citep{Karachentsev_2013} and their ${1.19^\circ}$ sky separation (corresponding to ${\ga 28}$ kpc) seem consistent with this scenario given that the distance to Antlia \citep[${1.31 \pm 0.03}$ Mpc,][]{Pimbblet_2012} is almost the same as that to NGC 3109 \citep[${1.286 \pm 0.015}$ Mpc,][]{Dalcanton_2009}. Thus, the galaxies are probably ${\la 40}$ kpc apart and may well be bound. Indeed, the latest results indicate that this is rather likely \citep[][Section 9]{Koruchi_2017}.

To account for this possibility, we exclude Antlia from our sample as it almost certainly has a much lower mass than NGC 3109 given that it is $\ssim 5$ magnitudes fainter \citep[][Table 3]{McConnachie_2012}. Although this hardly alters the properties of the plane best fitting the remaining HVGs (Table \ref{Plane_parameters}), the loss of a HVG increases the probability that the remaining ones end up close to a plane in a random MC trial. As a result, the proportion of MC trials meeting all our criteria more than triples but is still only ${5.2 \pm 0.4\perthousand}$. Even this rather small figure assumes that HIZSS 3 has an accurate distance measurement and so should be in our sample of HVGs. As discussed in Section \ref{No_HIZSS}, this may well be incorrect, further reducing the proportion of MC trials matching observations.

\subsection{Aspect ratio instead of physical thickness}
\label{Using_aspect_ratio}

So far, our results have been dominated by how likely it is to obtain a plane of HVGs with a physical thickness (Equation \ref{z_rms}) as small as observed. Another measure of anisotropy is the aspect ratio $A$ (Equation \ref{Aspect_ratio}). To see how sensitive our results are to which statistic is used, we repeat our nominal analysis using $A$ instead of $z_{rms}$, each time considering the combination of all but one HVG that yields a best-fitting plane with the smallest $A$. This very slightly increases the proportion of MC trials matching all three criteria, from ${1.48 \pm 0.10\perthousand}$ to ${1.62 \pm 0.01\perthousand}$. This is consistent with no change, suggesting that our results are not much dependent on which statistic is used to quantify the anisotropy of the HVGs.

\subsection{Using a flattened prior distribution}
\label{Flattened_prior}

Our results indicate that the HVGs are likely not distributed isotropically (Table \ref{Probability_table}). However, $\Lambda$CDM typically produces filamentary structures and sheets \citep[e.g.][]{Noh_2007}. This leads us to investigate whether a slightly flattened distribution of HVGs would be consistent with observations. For this purpose, we randomly\footnote{from an isotropic distribution} select a vector $\widehat{\bm{n}}$ and `flatten' the heliocentric position vectors $\bm{r}$ towards each mock HVG as well as to M31 and the centre of the MW using
\begin{eqnarray}
	\widetilde{\bm{r}} ~~=~~ \bm{r} ~-~ f\left( \bm{r} \cdot \widehat{\bm{n}} \right) \widehat{\bm{n}} 
	\label{z_compression}
\end{eqnarray}

\begin{figure}
	\centering 
		\includegraphics [width = 8.5cm] {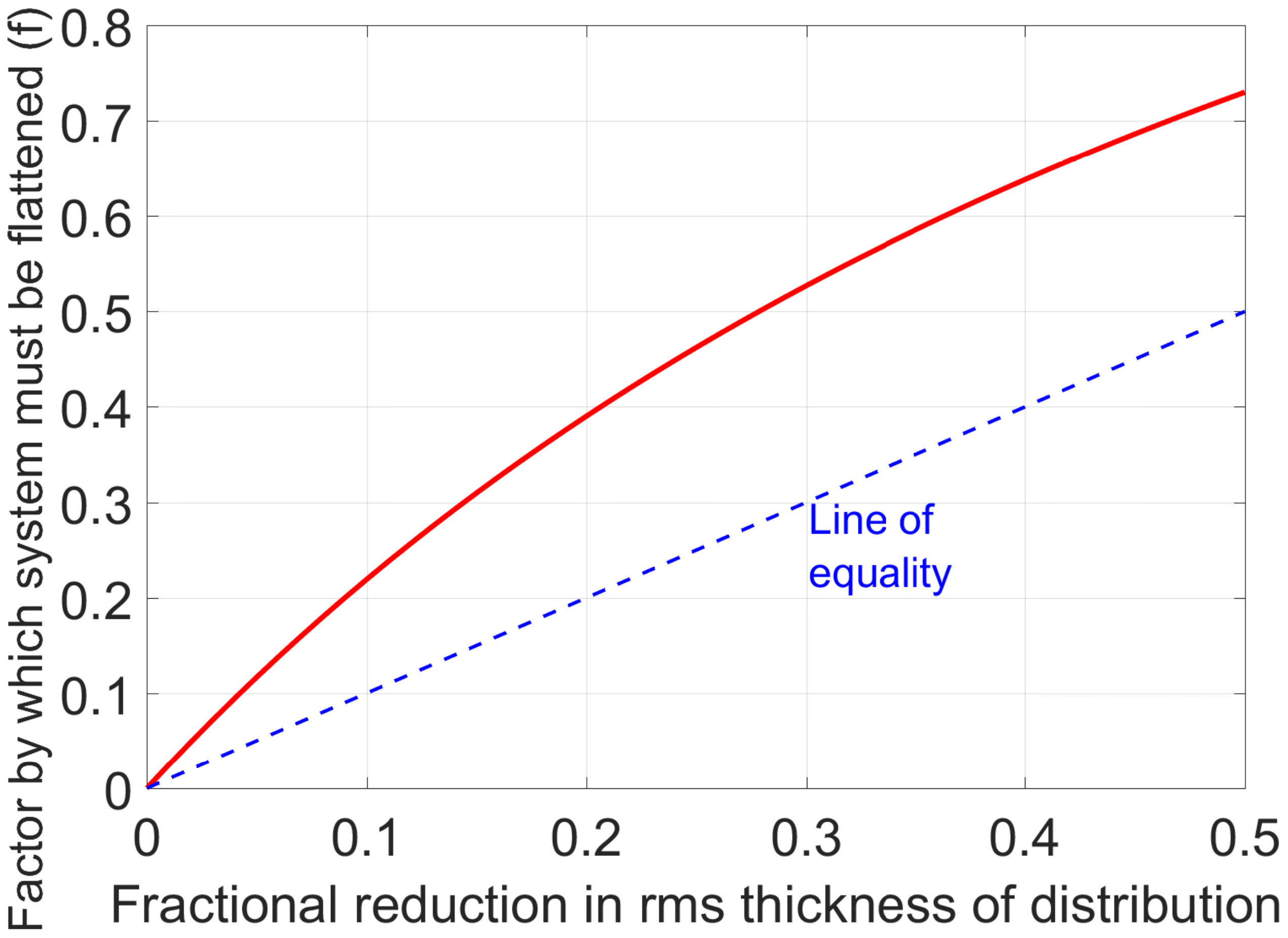}
	\caption{The solid red line shows the factor by which mock galaxy positions must be `flattened' ($f$ in Equation \ref{z_compression}) as a function of the desired fractional reduction in the rms thickness of an initially isotropic distribution of particles kept at fixed distances. This last requirement implies a non-trivial relation between the quantities plotted as any $f > 0$ generally implies a reduction in distance to an object, requiring a rescaling of its position vector (Equation \ref{Radial_rescaling}). As a result, the fractional reduction in rms thickness is much smaller than $f$. This is evident from our results lying left of the line of equality (dashed blue).}
	\label{Flattening_parameter_results_NR}
\end{figure}

A straightforward application of Equation \ref{z_compression} would put these objects at a smaller distance than observed. This is resolved by a subsequent rescaling of the position vector.
\begin{eqnarray}	
	\widetilde{\bm{r}} ~~\to ~~ \frac{\left| \bm{r} \right|}{\left| \widetilde{\bm{r}} \right|} ~\widetilde{\bm{r}}
	\label{Radial_rescaling}
\end{eqnarray}

This radial rescaling means that the final value of $z_{rms}$ is reduced by a smaller fraction than $f$. Taking an isotropically distributed set of unit vectors $\widehat{\bm{r}}$, the variance in the component of $\widetilde{\bm{r}}$ along $\widehat{\bm{n}}$ is
\begin{eqnarray}
	{z_{rms}}^2 &=& \int_0^\frac{\pi}{2} \frac{\alpha^2 \cos^2 \theta}{\alpha^2 \cos^2 \theta + \sin^2 \theta} \sin \theta ~d\theta \\
	&=& \frac{\alpha^2}{\beta^2} \left[ \frac{1}{2 \beta} Ln \left( \frac{1 + \beta}{1 - \beta}\right) - 1\right] ~\text{ where}\\
	\alpha &\equiv & 1 - f ~\text{ and}\\
	\beta &\equiv & \sqrt{1 - \alpha^2}
\end{eqnarray}

The integral can be solved by substituting for ${\beta \cos \theta}$. This lets us determine the value of $f$ required to obtain a desired fractional reduction in the $z_{rms}$ of an isotropically distributed unit vector via the successive application of Equations \ref{z_compression} and \ref{Radial_rescaling}. We achieve this using a Newton-Raphson root-finder. Our results are shown in Figure \ref{Flattening_parameter_results_NR}.

\begin{figure}
	\centering 
		\includegraphics [width = 8.5cm] {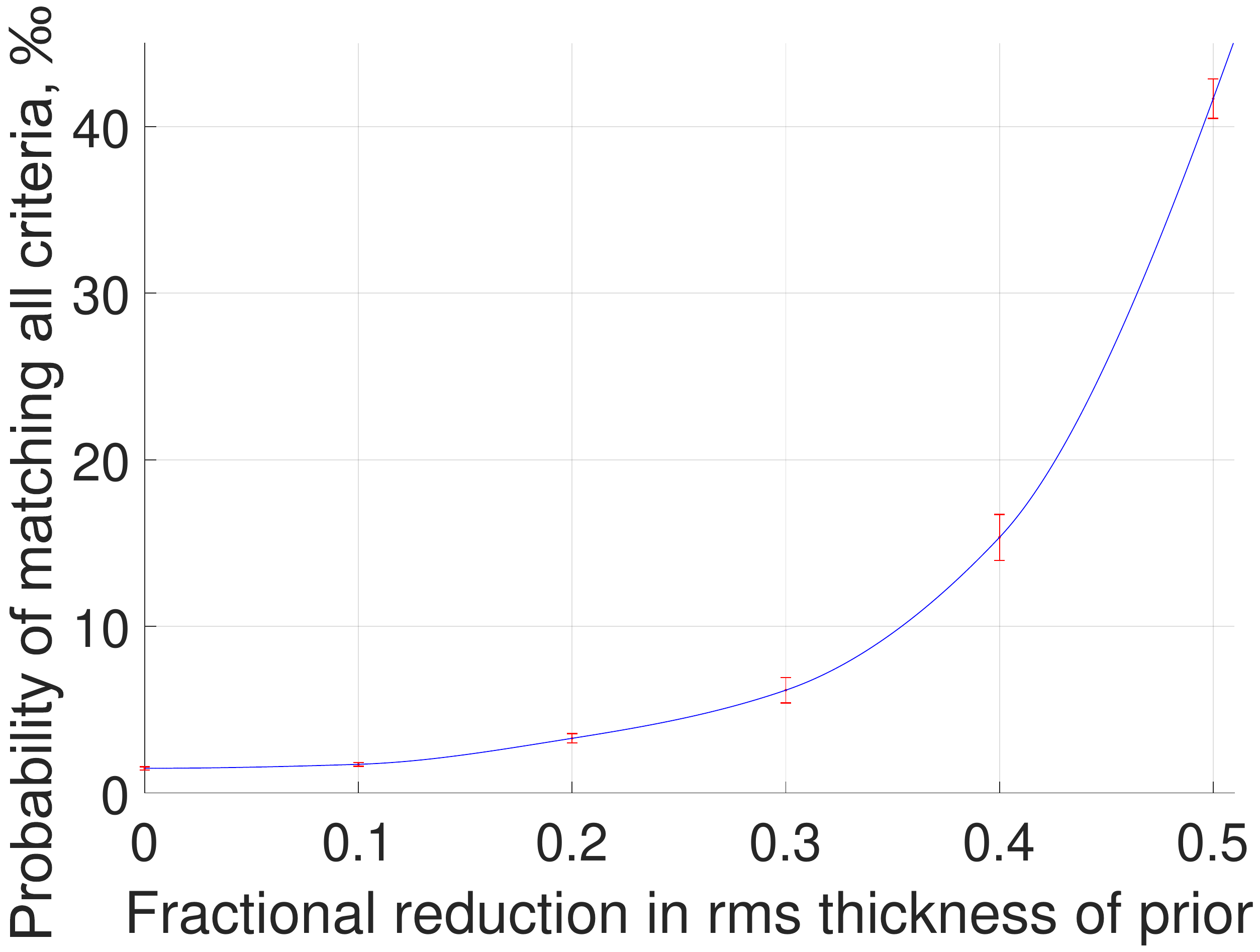}
	\caption{The effect on our MC analysis of flattening our prior for the spatial distribution of mock galaxies. The value of $f$ used in Equation \ref{z_compression} can be found using Figure \ref{Flattening_parameter_results_NR}. The probabilities shown are those of obtaining a mock catalogue of HVGs with properties analogous to the observed ones, according to the criteria defined in Table \ref{Criteria_definitions}.}
	\label{Flattening_results}
\end{figure}

In Figure \ref{Flattening_results}, we show the results of repeating our MC analysis using flattened prior distributions for the HVGs. We cover priors ranging from spherical to moderately oblate, similar to the range expected in $\Lambda$CDM for individual halos \citep{Butsky_2016}. Considering the positions of the galaxies in our full sample (not just HVGs), this seems a reasonable hypothesis for the LG (Figure \ref{Plane_offset_Delta_GRV_GA}). However, a mildly flattened underlying distribution is inconsistent with the observed HVG system as that has an aspect ratio $< 0.1$ (Table \ref{Plane_parameters}). Our results suggest that a flattening of ${\ga 0.5}$ would be required for consistency with observations.

To test whether a flattening of this sort is likely to be present in the LG, we apply our plane-fitting procedure (Section \ref{Best_fitting_plane_finding}) to our complete sample of LG galaxies, not just the HVGs. We find a preference for some flattening with respect to the axis $\left( 210^\circ,-6^\circ \right)$ in Galactic co-ordinates, though the rms thickness of the best-fitting plane is rather high at 645 kpc. The aspect ratio of our whole sample is 0.358, close to half the value of $\frac{1}{\sqrt{2}}$ that we would get for a purely isotropic distribution. If the underlying distribution of HVGs is flattened to this extent, then the probability of matching all criteria rises to ${41.7 \pm 1.2 \perthousand}$.

However, we also need to account for the ${25^\circ}$ mismatch between the axis about which the LG is flattened and the HVG plane normal. Although uncertainties of ${10^\circ}$ are possible in the latter, they should be much smaller in the former due to the larger number of galaxies (35 instead of 8). Thus, it is difficult to explain our results as a consequence of our entire sample being flattened $-$ it is, but about a different axis to the HVGs. Of course, our sample of LG galaxies may be subject to observational biases that make discovery more likely in certain directions on the sky. But the HVGs are likely subject to the same biases, unless there are selection effects based on the radial velocity.

To quantify whether this misalignment is natural, we add the requirement\footnote{to those listed in Table \ref{Criteria_definitions}} that the random `flattening axis' $\widehat{\bm{n}}$ in each MC trial should be misaligned by at least as much as the observed ${25^\circ}$ to the plane normal best-fitting the mock HVGs. Including this additional requirement decreases the proportion of analogues to just ${1.08 \pm 0.02 \perthousand}$. This is not very surprising $-$ if a MC trial `flattened' along $\widehat{\bm{n}}$ yields a thin plane of mock HVGs with rms thickness lower than the observed 101 kpc, it probably does so because the normal to the mock HVG plane aligns rather closely with $\widehat{\bm{n}}$.

Because we include the HVGs when finding the properties of our entire sample, any plane preferred by the remaining galaxies would get tilted towards the plane preferred by the HVGs. To account for this, we exclude the HVGs and fit a plane to the remaining galaxies. Surprisingly, this yields an even more pronounced flattening, with the aspect ratio decreasing to 0.24. However, this also increases the misalignment between the flattening direction and the HVG plane normal. To see how these competing effects alter our results, we repeat our analysis with a more flattened prior distribution but require $\widehat{\bm{n}}$ to misalign with the mock HVG plane normal by at least ${35^\circ}$. This makes analogues to the observed HVG system even harder to find, with their frequency falling to just ${0.12 \pm 0.01 \perthousand}$.


\section{Discussion}
\label{Discussion}

Based on the criteria in Table \ref{Criteria_definitions}, the observed system of HVGs appears most consistent with an underlying isotropic distribution if we use the aspect ratio (Equation \ref{Aspect_ratio}) instead of rms thickness (Equation \ref{z_rms}) and also remove Antlia from our sample while leaving HIZSS 3 in. Even with this highly favourable situation, only ${5.35 \pm 0.02 \perthousand}$ of the MC trials can be considered analogous to the observed situation (Table \ref{Probability_table}). Thus, it seems clear that an isotropic distribution of HVGs is very unlikely to mimic several aspects of the observed situation. This is a little unusual given that our full sample of galaxies does not prefer positions close to the plane defined by the HVGs (Figure \ref{Plane_offset_Delta_GRV_GA}).

However, the existence of HVGs and their anisotropic distribution are less puzzling in the context of our proposed scenario where the high GRVs arose through 3-body gravitational interactions of LG dwarfs with the MW and M31. Our simulation of this process (Section \ref{MOND_simulation}) implies that there should be a bimodal distribution of radial velocities at distances of ${\sim 1.5-3}$ Mpc from the LG. Observationally, this is hinted at in Figure \ref{Distance_GRV_correlation_3D_GA} of this work as well as in Figure 9 of both \citet{Banik_Zhao_2016} and \citet{Banik_Zhao_2017}.

The process is likely to be more efficient for a LG dwarf flung out parallel to the motion of the major LG galaxy it interacted with. Thus, the fastest HVGs $-$ which would now be furthest away from the LG $-$ should preferentially lie close to the MW-M31 orbital plane, which of course also includes the MW-M31 barycentre (Figure \ref{z_distribution}).

Even without detailed MC calculations, it is evident that most of the HVGs we identify (Table \ref{Planar_backsplash_galaxies}) lie rather close to a plane (Figure \ref{Plane_offset_Delta_GRV_GA}). However, this is not true of our sample in general. There are dozens of galaxies (the vast majority) whose kinematics are consistent with expectations based on $\Lambda$CDM. These galaxies neither avoid nor prefer the plane defined by the HVGs, suggesting that their unusual kinematics is related to their anisotropic spatial arrangement.

Looking at the results of our statistical calculations (Table \ref{Probability_table}) in a less model-dependent way, one might focus on the probability of obtaining a plane thinner than observed (i.e. using just the first or second criterion in Table \ref{Criteria_definitions} and ignoring the others). In this case, our results are ${\ssim 4 \times}$ larger. The highest result of ${22.2 \pm 0.1 \perthousand}$ arises when removing Antlia from our sample and using the aspect ratio instead of physical thickness.


The statistical significance of our results arises almost exclusively due to the galaxies in the NGC 3109 association, whose filamentary nature has been discussed previously \citep{Bellazzini_2013}. Looking at Figure 1 of that work, it is clear that its conclusions are strengthened by a more recent distance measurement to Leo P \citep[$1.62 \pm 0.15$ Mpc,][]{McQuinn_2015}. The rather high GRVs of these galaxies were noted by \citet{Teyssier_2012} and \citet{Shaya_2013}, the latter work finding no choice but to assume a past gravitational slingshot interaction with the MW ${\ssim 7}$ Gyr ago at a closest approach distance of 25 kpc (see their Section 4.2.4). Although we consider this very likely, the role of dynamical friction in such close encounters was neglected, possibly missing an opportunity to discriminate between $\Lambda$CDM and modified gravity alternatives \citep{Pawlowski_McGaugh_2014}. Section 4.2 of that work reviews various explanations for the origin of both the spatial anisotropy and the kinematics of this association. However, the work does not consider the whole LG and lacks detailed dynamical modelling.

It is important to test if a scenario similar to what we propose for the LG might also have occurred elsewhere in the Universe. However, it is difficult to perform a timing argument analysis outside the LG because distance uncertainties are larger, making it harder to accurately pin down deviations from the Hubble flow (Equation \ref{v_pec}). Nonetheless, the galaxy group containing NGC 1407 seems to have a much wider spread of radial velocities than is suggested by the variation in line of sight distances \citep{Tully_2015}. In particular, the galaxies NGC 1400 and NGC 1407 have a radial velocity difference of nearly 1200 km/s even though it is quite likely that they are within 10 Mpc of each other, possibly much less \citep{Tully_2013}. Several other galaxies in the group also show a rather wide spread in radial velocity despite contamination issues not being so severe \citep{Trentham_2006}. Although there is no clear evidence for a recent NGC 1400-NGC 1407 interaction \citep{Spolaor_2008}, the galaxies are rather gas-poor ellipticals that could not be expected to long retain (or ever have) obvious features of a close interaction e.g. a recent starburst or tidal tails. Recent radio outbursts in the general vicinity are suggestive of the cluster gas sloshing around due to past motion of massive objects \citep{Giacintucci_2012}. Deeper observations targeting more accurate distance measurements could help clarify the kinematics of this system.

\subsection{Understanding our results in MOND}
\label{Discussion_MOND}

In our scenario, the HVGs should preferentially lie near a particular plane with normal along $\widehat{ \bm h}$, the direction of the MW-M31 orbital angular momentum (Figure \ref{z_distribution}). However, given the fairly small number of HVGs in our sample, it is possible for the observationally determined best-fitting plane to differ from any true underlying plane that is statistically preferred by the HVGs. This is because our MOND-based simulation of the LG (Section \ref{MOND_simulation}) indicates that not all the HVGs should orbit within exactly the same plane (Figure \ref{Cos_angle_histogram_1_6_Mpc}).

Moreover, there must be observational errors in our determination of the HVG plane normal. Assuming distance errors of ${\ssim 0.1}$ Mpc for a structure ${\ssim 1}$ Mpc wide, this translates into an uncertainty of ${6^\circ}$, though the actual uncertainty is smaller as we have more than one HVG. There are also uncertainties in determining exactly which galaxies should be considered as HVGs. For example, removing Antlia from our sample tilts the HVG plane by ${2.5^\circ}$ (Table \ref{Plane_parameters}). This suggests observational errors of up to ${\ssim 10^\circ}$ in the HVG plane orientation.

Thus, we do not expect a perfect alignment between our estimated direction for $\widehat{\bm h}$ (Section \ref{Orbital_plane_finding}) and that which best fits our sample of HVGs (Table \ref{Plane_parameters}). Nonetheless, the ${19^\circ}$ angle between these may be a problem for our scenario. In particular, the fact that the MW-M31 line is ${16^\circ}$ out of the plane best fitting the HVGs could suggest that we are missing something. This angle is reduced to $15^\circ$ if we remove Antlia from our sample of HVGs (Section \ref{Excluding_Antlia}), but is unlikely to be reduced to 0 given expected observational uncertainties.

\begin{figure}
	\centering 
		\includegraphics [width = 8.5cm] {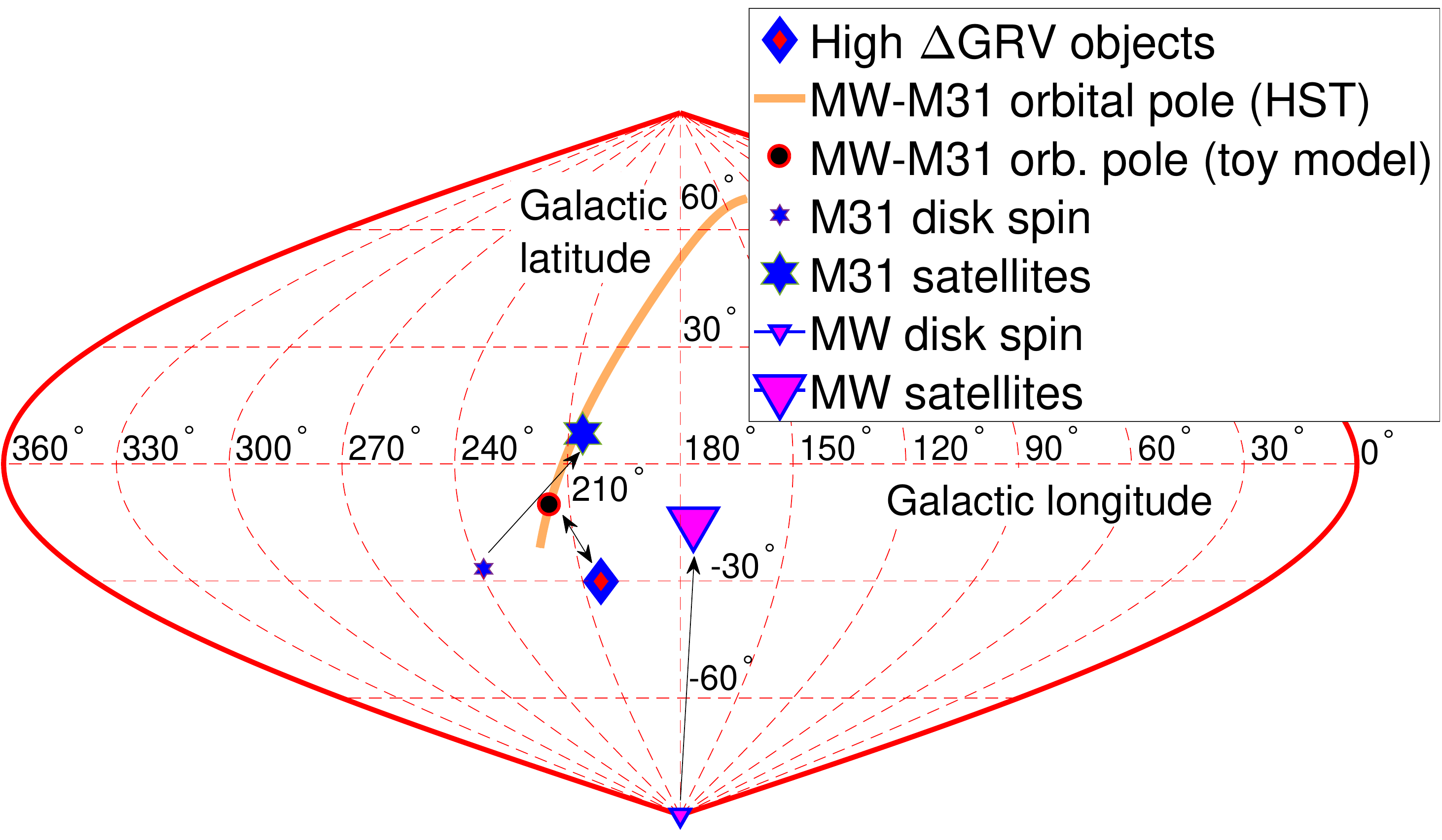}
	\caption{Directions of the vectors important to this work, in Galactic co-ordinates. Assuming a past close MW-M31 flyby, we expect tidal torque from M31 to explain the misalignment between the orientation of the MW disk (small triangle) and its plane of satellites (large triangle). The effect of such torques is illustrated using an upward arrow. Tidal torque from the MW explains a similar misalignment for M31 (hexagrams used instead of triangles). Our MOND-based toy model is able to reproduce these orientations fairly well if the MW-M31 orbital pole lies in the direction of the black dot with red rim (Section \ref{Orbital_plane_finding}). This is reasonably consistent with the normal to the plane defined by the HVGs (diamond), though we do not yet know its sense of rotation. The proper motion of M31 has recently been measured \citep{Van_der_Marel_2012}, suggesting a particular MW-M31 orbital pole (1$\sigma$ allowed region shown as orange line). This must be orthogonal to the present direction towards M31. Unfortunately, at 2$\sigma$, any direction consistent with this requirement is allowed.}
	\label{Directions_plotting_3D}
\end{figure}

However, there may well be deficiencies in our model which account for this discrepancy. One shortcoming is our treatment of the MW and M31 as point masses which have remained almost constant with time. We do not expect this to invalidate the main conclusion of our MOND model, namely that the particles flung out at high speeds by the MW/M31 tend to lie close to the MW-M31 orbital plane (Section \ref{MOND_simulation}). This is because encounters in MOND are similarly strong regardless of the impact parameter due to a cancellation between the encounter duration and the forces acting during the encounter. As a result, the vast majority of our simulated HVG analogues never passed within 5 disk scale lengths of either the MW or M31 (Figure \ref{MOND_simulation}) $-$ the commoner more distant encounters are strong enough.

These high-velocity particles gain most of their velocity in a cosmologically brief time period. Thus, our results should still hold if the MW and M31 masses are allowed to vary with time in a more realistic way. At present, it is unclear what a realistic MOND accretion history might look like. However, there is good reason to believe that structure formation would be more efficient than in $\Lambda$CDM \citep{Llinares_2008}. This would mean that galaxies drain their surroundings efficiently at rather early times, to some extent justifying our `island universe' approximation. Moreover, for a timing argument analysis, it is not important if gas 100 kpc from the MW gets accreted into its disk and triggers star formation. This would have significant consequences for the MW, but from the perspective of e.g. M31 at a distance of ${\ssim 800}$ kpc \citep{McConnachie_2012}, the gravitational attraction towards the MW would remain almost unchanged.

A potentially important effect beyond our point mass model is precession of the MW-M31 orbital pole ${\widehat{\bm{h}}}$ arising from the extended nature of their mass distributions and their close interaction. Due to the larger mass and disk scale length of M31, the more important consideration is how the orbit of the MW may have precessed due to the non-spherical nature of the M31 gravitational field. The general direction of this precession can be estimated by applying Equation \ref{Tidal_torque_direction} to the MW-M31 direction ${\bm{\widehat{r}}_{_{M31}}}$ at that time, which we find by rotating the present MW-M31 line about our best guess for ${\widehat{\bm{h}}}$ (Section \ref{Orbital_plane_finding}). This suggests that ${\bm{\widehat{r}}_{_{M31}}}$ lay along Galactic co-ordinates $\left( 270^\circ, 66^\circ \right)$ at closest approach. Obtaining the M31 disk spin vector from Table \ref{M31_image_parameters}, we find that ${\widehat{\bm{h}}}$ likely shifted towards the direction $\left( 334^\circ,-11^\circ \right)$ i.e. almost directly towards the orange line in Figure \ref{Directions_plotting_3D}, the locus of all directions where ${\widehat{\bm{h}}}$ could currently be given the direction towards M31. With a closest approach distance only a few times larger than the scale length of the M31 disk, ${\widehat{\bm{h}}}$ could conceivably have precessed by ${\ssim 10^\circ}$, enough to explain why ${\bm{\widehat{r}}_{_{M31}}}$ does not lie entirely within the HVG plane. More detailed models are required to account for such effects.

Another possibility is the effect of massive objects outside the LG. In \citet{Banik_Zhao_2016}, we found that Centaurus A has a discernible impact on LG dynamics in a $\Lambda$CDM context. The stronger long-range gravitational force in MOND may enhance such effects, especially as the non-linearity of the theory means that even a constant external gravitational field on the LG can influence its internal dynamics. In such circumstances, the force exerted on a test particle by a massive object might not be directed towards it \citep[e.g.][]{Banik_Zhao_2015}. Combined with tides, this can lead to tilting of a plane of LG galaxies defined by only a small number of them.

\begin{figure}
	\centering 
		\includegraphics [width = 8.5cm] {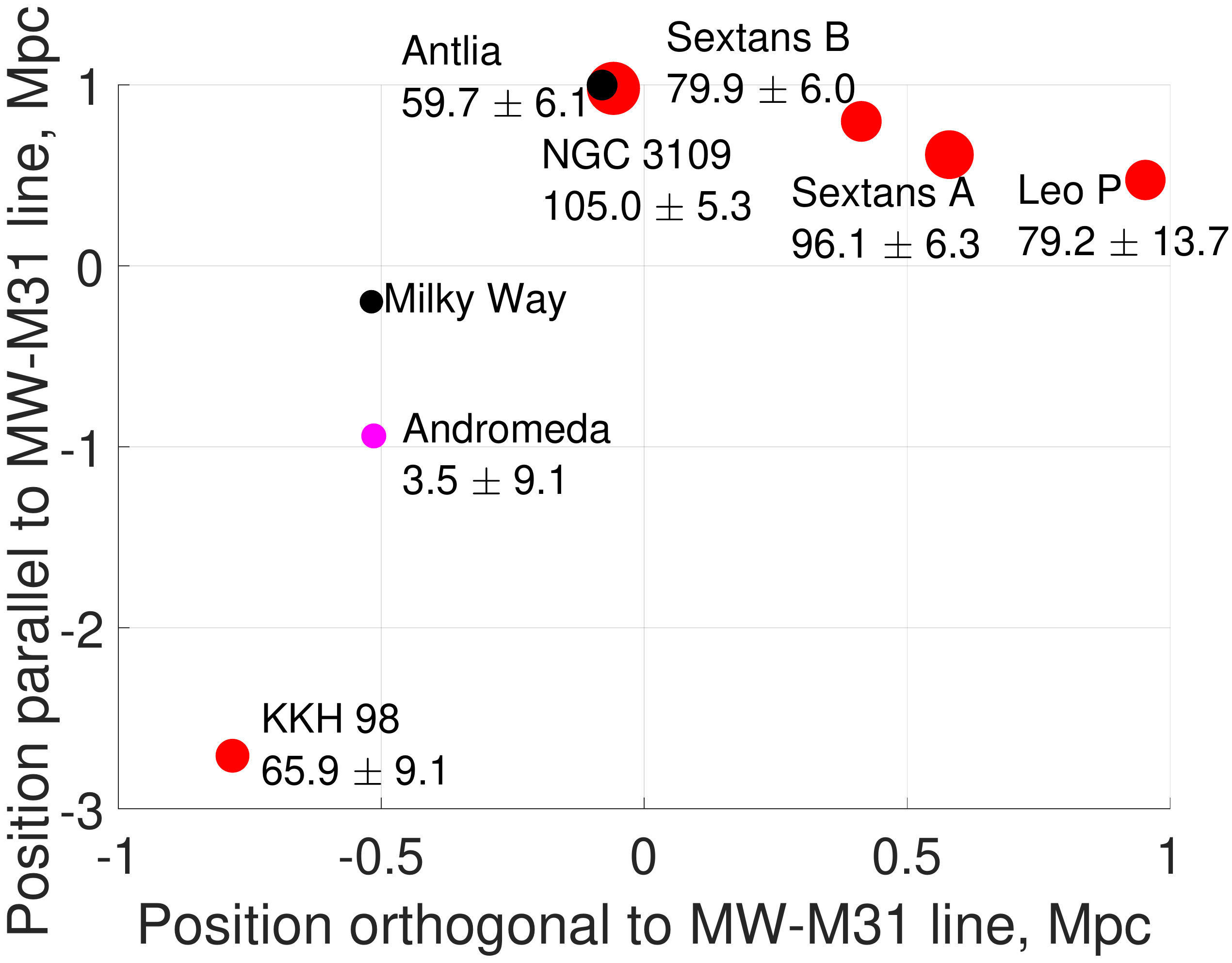}
	\caption{$\Delta GRV$s of indicated galaxies are shown against their position within the best-fitting plane through them (orientation given in Table \ref{Plane_parameters}). Only HVGs close to this plane are shown, with the marker size proportional to $\Delta GRV$ (except for the MW and M31). The number below the name of each galaxy also shows its $\Delta GRV$ in km/s. For clarity, position uncertainties are not shown.}
	\label{Plane_position_Delta_GRV_GA}
\end{figure}

Such effects can be enhanced by the filamentary configuration of the HVGs within the plane they seem to define (Figure \ref{Plane_position_Delta_GRV_GA}). Taking an extreme example, suppose a Cartesian $xyz$ co-ordinate system is used and that all the HVGs are located on the $xy$-plane along the lines ${L_1}$ and ${L_2}$ which are parallel to the $x$-axis. A perturber in the ${yz}$-plane at ${z \neq 0}$ would cause galaxies in ${L_1}$ to rise out of the ${xy}$-plane by a different amount to galaxies in ${L_2}$, such that the best-fitting plane through all the HVGs would no longer be the ${xy}$-plane. The analysis of such effects is beyond the scope of this investigation, but may be feasible based on purely geometric arguments.



\subsection{Understanding our results in $\Lambda$CDM}
\label{Discussion_LCDM}

It is possible that there is a $\Lambda$CDM-based explanation for the results discussed in this contribution. However, the anomalously high GRVs of the HVGs are unlikely to be explained by the tidal effects of large-scale structure \citep[][Section 5.1]{Banik_Zhao_2017} or by massive galaxy groups just outside the LG $-$ these are already directly included in our dynamical model (see its Table 3). Although that analysis could have missed dynamical solutions involving close interactions outside the LG, this is only likely to be a viable solution for distant galaxies like NGC 4163 which have an extremely low GRV compared to expectations (Figure \ref{Distance_GRV_correlation_3D_GA}). It is conceivable that this galaxy was flung towards the LG by a close interaction outside it which is not properly included in our simulation. However, there is insufficient time for a galaxy to be flung towards the LG and now be moving away from the LG again \citep[][Section 5.2]{Banik_Zhao_2017}. Moreover, our improvements to that analysis (Section \ref{Model_refinements}) make it less likely that we miss such encounters.

Our $\Lambda$CDM-based timing argument did not allow for the masses of galaxies changing with time, similar to previous works \citep[e.g.][]{Jorge_2014}. Although galaxies are expected to grow with time \citep[e.g.][]{Wechsler_2002}, this should not much affect our results for several reasons. The main one is that the timing argument is mostly sensitive to late times because an impulse at earlier times would change both the present position and velocity of a particle. This causes an effect similar to but steeper than Hubble drag \citep[][Figure 4]{Banik_Zhao_2016}. A reduction in masses at earlier times would also need to be compensated by a higher present mass in order to get a similar time-averaged gravitational field in the LG and thus to match observed GRVs. Moreover, for e.g. the MW-M31 gravitational attraction to be substantially affected by accretion on to the MW, the accreted material would need to come from a very large distance. Even if it came from just beyond the MW virial radius ${\ssim 200}$ kpc away \citep{Dehnen_2006}, this is still only a small fraction of the ${\ssim 800}$ kpc distance to M31. Thus, at the large scales investigated in this contribution, it should be acceptable to treat LG galaxies as having a fixed mass.

Given that our model handles tides raised by objects outside the LG and the paucity of galaxies with an unusually low $\Delta GRV$ (Figure \ref{Distance_GRV_correlation_3D_GA}), the most likely explanation for the HVGs is that they were flung out by massive fast-moving object(s) \emph{inside} the Local Group several Gyr ago. Our investigation does not elucidate the nature of these object(s), tempting as it may be to identify them with the MW and M31. This is infeasible in $\Lambda$CDM as the theory implies no past MW-M31 flyby $-$ in any close interaction, dynamical friction between their dark matter halos would cause a merger \citep{Privon_2013}. Without a past interaction, the MW and M31 would always have been slow-moving relative to the LG barycentre, limiting their ability to account for the HVGs.\footnote{This is not true at very early times, when the Hubble expansion was very fast. However, impulses at those times hardly affect present motions due to Hubble drag. We demonstrated this explicitly by showing that our results hardly changed when we started our simulations earlier \citep[][Section 4.6]{Banik_Zhao_2016}.}

Moreover, the growth of the MW and M31 masses with time implies that their scattering power must have been smaller at earlier times \citep{Wechsler_2002}. Indeed, MW and M31 analogues in $\Lambda$CDM simulations do not appear to fling out dark matter halos beyond $\ssim 3$ virial radii and even these `associated halos' have rather small outwards velocities \citep[][Figures 3 and 6]{Sales_2007}.

Dynamical friction is less efficient for less massive objects like M33 and the Large Magellanic Cloud, but both of these are already included in our model at velocities consistent with their observed GRVs and proper motions \citep[][Table 3]{Banik_Zhao_2017}. A galaxy with even less mass would have an even smaller circular velocity \citep{Evrard_2008}, making it unlikely that it could ever have scattered the HVGs onto their presently observed orbits. 

Given the significance of dynamical friction for heavier galaxies, it is possible that another galaxy $X$ merged with e.g. M31, building up a high velocity relative to it by falling deep into its potential well. Any LG dwarf passing close to the spacetime location of the $X$-M31 interaction could then be flung out at high speed, taking up some of the kinetic energy of $X$ if $X$ was sufficiently massive. However, too high a mass could disrupt the M31 disk and limit the $X$-M31 relative velocity due to more significant dynamical friction. On the other hand, too low a mass is also not feasible due to the need to scatter the HVGs we identified, some of which have rather high masses. For example, NGC 3109 is rotating at ${\ssim \frac{1}{3}}$ the rate of the MW so must have a few percent of its mass \citep{Jobin_1990}. Sextans A and B also have substantial masses of ${\ssim 10^9 M_\odot}$ \citep{Bellazzini_2014}. These considerations may leave a range of plausible masses for $X$, though we argued previously that the effects of Hubble drag imply it would only have enough scattering power to explain our results if its mass was comparable to that of the MW or M31 \citep[][Section 5.2]{Banik_Zhao_2017}.

In such a scenario, the past gravitational field in the LG would have been different to that assumed in our $\Lambda$CDM timing argument calculation. However, this is mostly sensitive to forces acting at late times, thus limiting the effect of an error in the gravitational field at early times \citep[][Figure 4]{Banik_Zhao_2017}. If we really have got the past gravitational field in the LG wrong at a recent enough time to affect our results, then one might expect some galaxies to have GRVs much below the predictions of our (erroneous) model. Our analysis indicates that this very rarely happens (Figure \ref{Plane_offset_Delta_GRV_GA}). Moreover, it is difficult to find analogues to the HVGs in cosmological simulations of $\Lambda$CDM that include realistic merger histories for analogues of the MW and M31 \citep[][Figure 2]{Sales_2007}.



A related scenario is that the HVGs were formed by tidal disruption of $X$, making them tidal dwarf galaxies flung out during the merger of a gas-rich galaxy $X$ with a major LG galaxy. The chaotic motions and complicated gas hydrodynamics during this interaction may provide a way around the fact that the slow motion of the MW and the even slower motion of M31 do not readily provide a mechanism to create HVGs. One consequence of this scenario is that the HVGs should have rather low internal velocity dispersions because tidal dwarf galaxies should be almost free of dark matter \citep{Wetzstein_2007}. However, the internal dynamics of Tucana \citep{Fraternali_2009} and NGC 3109 \citep{Jobin_1990} indicate that they require dark matter for dynamical stability in a $\Lambda$CDM context. This is also the case for Sextans A and B \citep{Bellazzini_2014}.

The dark matter in the HVGs could be understood if $X$ had its own retinue of satellite galaxies. The $X$-M31 interaction would disrupt this satellite system, perhaps creating a few HVGs. In one such scenario, $X$ is identified with NGC 205 and its satellites have now formed a structure analogous to a tidal stream around M31, helping to explain its anisotropic distribution of satellites \citep{Angus_2016}.

Assuming that some satellites of NGC 205 could escape from M31, an obvious consequence of this scenario should be that the HVGs lie in the same plane as that preferred by satellites of M31. However, there is a ${41^\circ}$ mismatch between the orientations of these planes (Figure \ref{Directions_plotting_3D}), far larger than likely observational uncertainties. The plane of M31 satellites may have precessed from its original orientation \citep{Fernando_2016}, but such a large amount of precession would almost certainly inflate the thickness of the structure as not all M31 satellites would be equally affected.



Although satellites of NGC 205 could plausibly end up at the fairly low velocities typical of M31 satellites, it is questionable whether they could become HVGs. Using a test particle cloud around NGC 205 to represent its satellites, it is clear that only a minuscule fraction (if any) of these particles end up further than 200 kpc from M31 at the present time \citep[][Figure 12]{Angus_2016}. This is because a fairly low infall velocity is required to ensure that a substantial number of NGC 205 satellites become bound to M31.

As the mass of NGC 205 which worked best in these models was only 1\% that of M31 itself \citep[][Section 2.2.2]{Angus_2016}, any satellites of NGC 205 would likely end up in a bound orbit around M31, just like NGC 205 itself. Applying the $M \propto {v_{_f}}^3$ scaling typical of $\Lambda$CDM halos \citep{Evrard_2008} and assuming $v_{_f} = 225$ km/s for M31 \citep{Carignan_2006}, we get ${\ssim 45}$ km/s for the typical velocity of a NGC 205 satellite relative to its host. This is much smaller than the difference between the circular and escape velocities of M31, which must be at least ${\left( \sqrt{2} - 1\right)} v_{_f} \approx$ 90 km/s. The actual figure could be far higher due to the extended nature of the M31 mass distribution. For example, the MW escape velocity 50 kpc from it is ${\ssim 380}$ km/s \citep{Williams_2017} despite ${v_{_f}}$ only being ${\ssim 200}$ km/s for the MW \citep{Kafle_2012}. Thus, it is easy to see why this scenario does not lead to a satellite of NGC 205 ending up as far from M31 as the HVGs.

We should bear in mind that this proposal was not designed to explain our HVGs. For this purpose, a variant could be considered where $X$ is no longer identified with NGC 205 but instead fell towards M31 with a much higher relative velocity. This might well lead to a filamentary structure receding from the LG at high speed, with $X$ perhaps identified as the NGC 3109 association. However, this does not explain the high infall velocity of $X$, which seems difficult to reconcile with $\Lambda$CDM \citep[][Section 5.2]{Banik_Zhao_2017}.

Moreover, a fairly massive dark matter halo would be needed to hold the NGC 3109 association together \citep{Bellazzini_2013}. It would have to pass quite close to the MW/M31 in order to get tidally disrupted, as would be required for the galaxies in this association to end up in their observed filamentary configuration. It is unclear whether dynamical friction would render such a scenario infeasible, though that appears quite likely if the association has a width of 600 kpc and a mass of ${3.2 \times 10^{11} M_\odot}$ \citep[as suggested by][]{Bellazzini_2013}. To account for the rather high GRVs of the galaxies in this association, it needs to have passed within 25 kpc of the MW \citep[][Section 4.2.4]{Shaya_2013} without subsequently merging or being significantly decelerated.

In addition to this issue, several other difficulties with the scenario were discussed in Section 4.2.1 of \citet{Pawlowski_2014}. Moreover, the latest results indicate that the galaxies in this association are very unlikely to be gravitationally bound and probably lie within their own individual dark matter halos \citep[][Section 9]{Koruchi_2017}. The only exception is Antlia, which could be bound to NGC 3109. However, excluding Antlia from our HVG sample does not greatly alter our conclusions (Section \ref{Excluding_Antlia}). If the remaining HVGs are not gravitationally bound to each other, then it is reasonable to consider these dwarf galaxies as independent. In this case, our results show that they have a statistically significant tendency to lie close to a plane (Table \ref{Probability_table}). Thus, it remains difficult to simultaneously explain the high radial velocities of the HVGs and their anisotropic spatial distribution within the $\Lambda$CDM paradigm.

\section{Conclusions}
\label{Conclusions}

We recently identified several Local Group (LG) galaxies with much higher radial velocities than can easily be understood in the context of $\Lambda$CDM \citep{Banik_Zhao_2016, Banik_Zhao_2017}. These high velocity galaxies (HVGs) are not bound to the Milky Way (MW) or M31 but instead lie ${\ga 1}$ Mpc away from them. This is ${\ga 5 \times}$ the MW virial radius \citep{Dehnen_2006}, well beyond the furthest distance to which cosmological simulations indicate that the MW or M31 could slingshot out dark matter halos containing LG dwarf galaxies \citep[][Figures 3 and 6]{Sales_2007}. The issue therefore arises on a much larger scale than that addressed in previous works regarding the anisotropic distribution of the MW and M31 satellite systems \citep[e.g.][]{Pawlowski_2014}. In a $\Lambda$CDM context, baryonic physics does not seem to have a major effect on the expected anisotropy of these systems \citep[][Figure 3]{Pawlowski_2015}. Thus, it seems unlikely that it would affect our results on a still larger scale.

In this contribution, we assess the feasibility of a scenario involving a past close flyby of the MW and M31, whose once fast relative motion could have flung out these HVGs via gravitational slingshot encounters. To help constrain the orientation of the MW-M31 orbital plane, we develop a toy model of their flyby forming the recently discovered planes of satellites around the MW \citep{Kroupa_2013} and M31 \citep{Ibata_2013}. A past close MW-M31 encounter seems able to form correctly oriented satellite planes only for a particular MW-M31 orbital pole (Section \ref{Orbital_plane_finding}).

Using this information, we constructed a MOND-based dynamical model of the LG to investigate the effect of a past MW-M31 flyby on it (Section \ref{MOND_simulation}). We tracked the evolution of a spherical cloud of several hundred thousand LG test particles initially on the Hubble flow, leaving a gap around the MW and M31 (Equation \ref{r_exc}). Although our results are somewhat dependent on details of how MOND works in a cosmological context, it is clear that a small fraction of these particles end up with a large radial velocity away from the LG after closely interacting with the MW or M31 around the time of their encounter (Figure \ref{Hubble_diagram_coloured_FL}). Such slingshot interactions are expected to be most efficient for particles flung out almost parallel to the motion of the perturbing body. This probably explains why simulated particles flung out to the greatest distances from the LG preferentially lie close to the MW-M31 orbital plane (Figure \ref{z_distribution}).

To see if such a pattern is evident in the real LG, we made several improvements to our $\Lambda$CDM modelling of the LG (Section \ref{Model_refinements}) and developed a method for selecting HVGs (Section \ref{Sample_selection}) so that we could quantify their spatial anisotropy (Section \ref{Method}). The HVGs we identified are indeed mostly located close to a plane (Figure \ref{Plane_offset_Delta_GRV_GA}) oriented similarly to our expectation for the MW-M31 orbital plane based on the flyby scenario (Figure \ref{Directions_plotting_3D}). The galaxies in our sample which are not HVGs (the vast majority) do not preferentially lie close to the HVG plane. Importantly, the MW and M31 lie near this plane (barycentre $\ssim 70$ kpc off it), suggesting that it has dynamical significance.

These characteristics of the observed HVG plane are natural outcomes of our MOND simulation of the LG (Section \ref{MOND_simulation}). To quantify how likely it is that a random distribution of HVGs shares these properties, we conducted Monte Carlo (MC) trials where the directions to the HVGs were randomised and their distances selected from a Gaussian distribution corresponding to observations (Section \ref{Results}).\footnote{We converted negative mock distances to 0.}

Based on the criteria in Table \ref{Criteria_definitions}, there is a ${< 0.01}$ probability of obtaining a situation analogous to that observed, mainly because it is very unusual for all but one of the HVGs (listed in Table \ref{Planar_backsplash_galaxies}) to so nearly lie within a common plane. Our result holds even if we vary certain assumptions, with the nominal choices leading to a still smaller value of only ${1.5 \pm 0.1 \perthousand}$ (Table \ref{Probability_table}). As the observed HVG plane has an aspect ratio of 0.08 (Table \ref{Plane_parameters}), our results are not consistent with a mildly flattened prior distribution for the HVGs (Section \ref{Flattened_prior}). More extreme flattening could lead to consistency and may well be realistic as our entire sample of LG galaxies does indeed exhibit a moderate flattening. However, the preferred flattening axis is misaligned by ${25^\circ}$ \footnote{${35^\circ}$ when excluding HVGs} to the HVG plane normal. Thus, the flattened distribution of the HVGs is very likely related to their anomalous radial velocities and not merely a consequence of the entire LG being moderately anisotropic about a different axis.



As well as more detailed modelling, our ideas regarding the LG may be tested using a more accurate distance to HIZSS 3 to see if it really should be treated as a HVG. Better observations of Antlia could help clarify whether it is a satellite of NGC 3109 or an independent galaxy. In the long run, it is important to try and discover more HVGs to see if they mostly remain close to a plane.\footnote{Observers do not search for HVGs per se but instead for LG dwarf galaxies which may turn out to be HVGs when their kinematics are analysed using a $\Lambda$CDM-based dynamical model.} If so, it will be interesting to know whether its orientation is altered much compared to our determination. Considering the effect of Antlia on this, adjustments of up to $\ssim 5^\circ$ could well be in store (Table \ref{Plane_parameters}).

Our work suggests a particular great circle on the sky in which observers would be more likely to discover HVGs. However, it is important to search other directions too as only then is it possible to determine just how anisotropic the distribution of HVGs is. Fortunately, the plane we identified is inclined by a large angle to the MW disk and to the Ecliptic, which should help minimise observational biases. Moreover, there are unlikely to be any strong selection effects based on the radial velocity.

Both 2D and 3D $\Lambda$CDM-based dynamical models of the LG face difficulties in explaining the observed kinematics of its non-satellite galaxies \citep{Banik_Zhao_2016, Banik_Zhao_2017}. In particular, the existence of 7 galaxies with anomalously high radial velocities but only 1 with anomalously low radial velocity (Figure \ref{Distance_GRV_correlation_3D_GA}) suggests that the central region of the LG has been much more efficient at scattering dwarf galaxies than these models allow. This is problematic given that even LG galaxies as massive as the MW and M31 appear incapable of scattering dwarfs as far out as the HVGs, not only in our simulations but also when considered in full cosmological simulations of $\Lambda$CDM \citep{Sales_2007}. This may hint that the MW and M31 were once moving much faster than at present, pointing towards a past close MW-M31 flyby. If such an event occurred, it would provide a natural explanation for several aspects of the spatial distribution of the HVGs, especially their tendency to lie close to a plane (Figures \ref{z_distribution} and \ref{Plane_offset_Delta_GRV_GA}). A past flyby interaction of such massive galaxies only makes sense in the context of certain modified gravity theories where galaxies lack massive dark matter halos and their associated dynamical friction in close encounters, which would otherwise cause a merger.


\section{Acknowledgements}
\label{Acknowledgements}

The authors are grateful to Marcel Pawlowski for suggesting the idea behind this contribution. IB is supported by Science and Technology Facilities Council studentship 1506672. This work was partly supported by a Scottish Universities Physics Alliance travel award hosted by Nima Arkani-Hamed at the Institute for Advanced Studies. The algorithms were set up using \textsc{matlab}$^\text{\textregistered}$.

\bibliographystyle{mnras}
\bibliography{LGA_bbl}
\bsp
\label{lastpage}
\end{document}